\documentclass[12pt]{article}%
\usepackage{subcaption}
\usepackage{adjustbox}
\usepackage{bm}
\usepackage{amsfonts}
\usepackage[colorlinks,citecolor=blue,urlcolor=blue,hyperfootnotes=false]%
{hyperref}
\usepackage{amssymb}
\usepackage[bottom]{footmisc}
\usepackage[hyperref]{ntheorem}
\usepackage{natbib}
\usepackage{graphicx}
\usepackage{float}
\usepackage[onehalfspacing, nodisplayskipstretch]{setspace}
\usepackage[margin=1in, a4paper]{geometry}
\usepackage{booktabs}
\usepackage{threeparttable}
\usepackage{amsmath}
\usepackage{lscape}
\usepackage{scrextend}
\usepackage{relsize}
\usepackage{multicol}
\usepackage{chngcntr}
\usepackage{etoolbox}
\usepackage{amssymb}
\usepackage{tabularx}
\usepackage[english]{babel}
\usepackage{multirow}
\usepackage{booktabs}
\usepackage[labelfont=bf]{caption}
\usepackage{float}
\usepackage{titlesec}
\usepackage{array}
\usepackage{color}
\usepackage{afterpage}
\usepackage{framed}
\usepackage{float}
\usepackage{afterpage}%
\usepackage{xr}
\providecommand{\U}[1]{\protect\rule{.1in}{.1in}}
\allowdisplaybreaks

\theoremstyle{plain}

\newtheorem{theorem}{Theorem}

\newtheorem{lemma}{Lemma}

\newtheorem{proposition}{Proposition}
\newtheorem{remark}{Remark}

\newtheorem{assumption}{Assumption}

\theorembodyfont{\normalfont}

\deffootnote{1em}{1.6em}{\thefootnotemark \enskip}

\numberwithin{equation}{section}
\numberwithin{theorem}{section}
\numberwithin{assumption}{section}
\numberwithin{proposition}{section}
\numberwithin{lemma}{section}

\makeatletter
\newcommand*{\addFileDependency}[1]{
  \typeout{(#1)}
  \@addtofilelist{#1}
  \IfFileExists{#1}{}{\typeout{No file #1.}}
}
\makeatother

\newcommand*{\myexternaldocument}[1]{%
    \externaldocument{#1}
    \addFileDependency{#1.tex}
    \addFileDependency{#1.aux}
}

\myexternaldocument{Appendix}

\begin{document}

\title{Specification tests for generalized propensity scores using double projections\footnote{We thank the Co-Editor (Professor Arthur Lewbel) and three referees for their valuable suggestions and comments.}}
\author{Pedro H. C. Sant'Anna\thanks{Emory University and Causal Solutions.
Email: pedrosantanna@causal-solutions.com}
\and Xiaojun Song\thanks{Guanghua School of Management and Center for Statistical Science, Peking
University. Email: sxj@gsm.pku.edu.cn. The author
acknowledges the financial support from the National Natural Science Foundation of
China (Grant No. 71973005).}}
\maketitle

\begin{abstract}
This paper proposes a new class of nonparametric tests for the correct specification of models based on conditional moment restrictions, paying particular attention to generalized propensity score models. The test procedure is based on two different projection arguments, leading to test statistics that are suitable to setups with many covariates, and are (asymptotically) invariant to the estimation method used to estimate the nuisance parameters. We show that our proposed tests are able to detect a broad class of local alternatives converging to the null at the usual parametric rate and illustrate its attractive power properties via simulations. We also extend our proposal to test parametric or semiparametric single-index-type models.
\end{abstract}

\thispagestyle{empty}

\noindent\textbf{Keywords}: Double projections; generalized propensity scores; multiplier bootstrap; single/multiple-index models; specification tests

\newpage\pagenumbering{arabic} \setcounter{page}{1} \setcounter{footnote}{0}

\section{Introduction}

One of the primary goals of many scientific fields is to quantify the effect of an exposure, policy, or treatment on outcomes of interest. When the assignment to treatment is not randomized, groups with different levels of the treatment variable usually differ in important ways other than the observed treatment. Because these differences are many times associated with the outcome variable, ascertaining the causal effect of the treatment requires more sophisticated statistical tools than a simple comparison of means. It is in this setting that the propensity score  and its multi-valued generalizations have been shown to be among the most widely used tools for causal inference; see, e.g.,  \cite{Imbens2015}, \cite{Linden2016} and \cite{Lopez2017}.

Although statistical procedures that build on the propensity score and its generalizations (henceforth GPS) are popular, a main concern of these methods is that the GPS is usually unknown, and therefore has to be estimated. Given the availability of many pre-treatment covariates and limited sample size, researchers usually adopt parametric models for the GPS. Such a common practice raises the important issue of model misspecifications. Indeed, as illustrated by \citet{Busso2014} and \cite{Linden2016}, model misspecifications can lead to misleading treatment effect estimates. Thus, in practice, assessing if your parametric putative model for the GPS is correctly specified is recommended. 

In this paper, we propose new goodness-of-fit tests for models based on conditional moment restrictions. Although our tests can be used more generally, we pay particular attention to testing whether a GPS model is correctly specified. The main distinguishing feature of our tests is that they combine two different projections. The first projection is a dimension-reduction device that allows us to handle situations with many covariates (\citealp{Escanciano2006b}). The second projection is used to make the test asymptotically invariant to the estimation method used to estimate the nuisance parameters (\citealp{Neyman1959, Bickel2006, Escanciano2014a, SantAnna2019}). As a result of this second projection, we can consider a wider range of estimators for nuisance parameters and implement a fairly simple bootstrap procedure that works for discrete responses, continuous responses, or a mixture of both.

This paper is the first to combine the two aforementioned projections. \cite{Escanciano2006b}'s test, although robust to dimensionality, is not robust to the estimation of nuisance parameters. A potential drawback of the test of \cite{Escanciano2006b} is that its implementation relies on the wild bootstrap, and it is not directly applicable when the response variables are discrete or mixed, which is the case in GPS models. On the other hand, the test of \cite{Escanciano2014a}, although robust to the estimation of nuisance parameters, is not robust to the presence of many covariates. The test of \cite{SantAnna2019}, although also constructed to be robust to many covariates and invariant to the estimation of nuisance parameters, only tests for an \emph{implication} of the null hypothesis, not for the null hypothesis itself. Furthermore, it is only justified for binary choice models. By contrast, our proposed tests are based on testing the null hypothesis itself, are robust to both the dimensionality of covariates and the estimation of nuisance parameters, and can be directly used for generic outcome models. 

Overall, our proposed tests enjoy several attractive features. They (a) do not severely suffer from the ``curse of dimensionality'' when we have many covariates; (b) are data-driven and do not require tuning parameters such as bandwidths; (c) do not require estimators to be $n^{1/2}$-asymptotically linear, with $n$ the sample size; and (d) are able to detect a broad class of local alternatives converging to the null at the parametric rate. In order to facilitate its practical implementation, we obtain closed-form expressions for our test statistics and show that critical values can be computed with the assistance of an easy-to-use multiplier-type bootstrap. To the best of our knowledge, no other (specification) test available in the literature enjoys all these attractive properties (e.g., \citealp{Escanciano2006b}, \citealp{Mora2008}, \citealp{Shaikh2009}, \citealp{Escanciano2014a}, \citealp{Garcia-Portugues2014}, \citealp{iwasawa2015joint}, \citealp{Zhu2017}, \citealp{SantAnna2019}, \citet{dominguez2020specification}, and \citealp{Kim2020}). The results of Monte Carlo simulations indicate that these attractive properties translate to tests with excellent finite sample properties, even when the dimension of covariates is relatively large.

We also consider extensions of our basic setup to other more complex frameworks. For instance, practitioners routinely use parametric GPS models that incorporate index restrictions. Single-index models are popular with binary or ordered multinomial treatments. With more general unordered, multinomial choice models, practitioners often use multiple-index models such as multinomial logit/probit. If such specifications were to be rejected by our omnibus tests, it would not be clear if it is due to violations of the index restriction, the parametric distributional assumption, or both. We extend our proposal to test index models to shed light on these concerns. Using our double-projection arguments, we propose a test for a parametric index model against semiparametric alternatives that maintain the index restrictions (directional test). We also extend our framework to test a single-index GPS model with an unknown link function against general nonparametric alternatives. We highlight that our double-projection procedure can eliminate the \emph{parametric}-type estimation effect in this semiparametric setup, though it does not eliminate the \emph{nonparametric}-type effect coming from estimating the link function. This additional complication suggests that one should use a different, much more demanding multiplier-bootstrap procedure to compute critical values. Although we discuss such challenges, we believe that a detailed solution to these issues is beyond the scope of this paper, and we leave it for future research.

The rest of this article is organized as follows. In Section \ref{DC}, we present the testing framework and introduce our proposed double-projection specification tests. The asymptotic properties of our tests are established in Section \ref{asy}, with the asymptotic null distribution derived in Section \ref{asy_null} and the asymptotic power properties studied in Section \ref{asy_alter}. In Section \ref{boot}, we present a detailed description of an easy-to-implement multiplier-bootstrap procedure to compute critical values. In Section \ref{extension_sim}, we extend our testing procedure to test index-type models; Section \ref{extension_sim1} tests a parametric index model against a nonparametric alternative, Section \ref{extension_sim2} tests a parametric index model against a semiparametric alternative, and Section \ref{extension_sim3} tests a semiparametric single-index model against a nonparametric alternative. We then illustrate the usefulness of our tests through an empirical application in Section \ref{empirical_section}. Section \ref{conclusion} concludes. Mathematical proofs and Monte Carlo simulation results are gathered in an online supplementary appendix.\footnote{Appendix is available at \url{psantanna.com/files/GPS_Appendix.pdf}}

\section{Specification tests based on double projections\label{DC}}

\subsection{Setup and motivation\label{setup}}

In this paper, we seek to test hypotheses of the type of
\begin{equation}
H_{0}:\mathbb{P}\left(  \mathbb{E}\left[  e(t;\theta_0)|X\right] =0\right)  =1\text{ for some }\theta_0\in\Theta\subset\mathbb{R}^{d_\theta}\text{ and all }t\in\mathcal{T}, \label{cmr}%
\end{equation}
against
\begin{equation}
H_{1}:\mathbb{P}\left(  \mathbb{E}\left[  e(t;\theta)|X\right]  =0\right)<1\text{ for any }\theta\in\Theta\subset\mathbb{R}^{d_\theta}\text{ and/or some } t\in\mathcal{T}, \label{cmr_alter}%
\end{equation}
where $e(t;\theta)$ is a generalized error term indexed by $t \in \mathcal{T} \subset \mathbb{N}$, and $\theta \in  \Theta\subset\mathbb{R}^{d_\theta}$, with $\Theta$ a compact space with $d_\theta\geq 1$ a given positive integer. For notational simplicity, we suppress the dependence on $t$ in $\theta$ and related quantities. Clearly, the null hypothesis $H_0$ is composite. Our main goal is to propose tests of $H_{0}$ against $H_{1}$ that are robust to the presence of many covariates\footnote{In this paper, ``many covariates'' means that the dimension $d_x$ of the covariates $X$ is allowed to be large but finite. For the derivation of our asymptotic results, we restrict $d_x$ to be fixed.} and the estimation of nuisance parameter $\theta_0$.

Although our framework is general, it is worth motivating within a popular and empirically relevant causal inference setup. For a generic $d_{1}\times d_{2}$ matrix $A$, $A^{\top}$ denotes the transpose of $A$. Let $J\in\mathbb{N}$ be a given finite positive integer. Let $(X^{\top},T,Y)^{\top}$ be a random vector in a $(d_{x}+2)$-dimensional Euclidean space, where $X\in$ $\mathcal{X}\subseteq\mathbb{R}^{d_{x}}$ is an observable $d_{x}\times1$ vector of covariates with $d_{x}\in\mathbb{N}$, $T$ $\in\mathcal{T\subseteq}\{0,1,\ldots,J\}$ is the treatment random variable, $Y=\sum_{t=0}^{\mathcal{T}}1\left(  T=t\right)  Y\left(  t\right) \in\mathcal{Y}\subseteq\mathbb{R}$ is the observed outcome, and $Y\left(t\right)  $ denotes the potential outcome when $T$ is externally set to $t$. For the sake of simplicity, let's focus on average causal effects of the form
\begin{equation}
\beta(t,s)\equiv\mathbb{E}\left[  Y\left(  t\right)  -Y\left(  s\right) \right]  , \label{ATE_def}%
\end{equation}
the average causal effect of exposing units to treatment $t$ rather than treatment $s$. We can also cover quantile treatment effects and/or treatment effects for the treated subpopulation (\citealp{Lee2018a}, \citealp{Ao2019}).

\cite{Imbens2000} shows that, if assignment to treatment $T$ is weakly unconfounded given the pre-treatment variables $X$, $\beta(t,s)$ in (\ref{ATE_def})  is identified by the following weighting estimands:%
\begin{equation}
\beta(t,s)=\mathbb{E}\left[  \frac{Y1\left(  T=t\right)  }{p_{t}\left(  X\right) }\right]  -\mathbb{E}\left[  \frac{Y1\left(  T=s\right)  }{p_{s}\left( X\right)  }\right], \label{ATE}%
\end{equation}
where $p_{t}\left(  x\right)  \equiv \mathbb{P}\left(  T=t|X=x\right)  =\mathbb{E}\left[  1\left(  T=t\right)
|X=x\right]  $ is the unknown GPS. To estimate $\beta(t,s)$ using (\ref{ATE}), researchers usually assume a parametric model $q_{t}\left(  x,\theta_{t}\right)$ for $p_{t}\left(  x\right)$, where $q_{t}(x,\theta_{t}):\mathcal{X}\times\Theta_t\mapsto\lbrack0,1]$ denotes a family of parametric functions known up to the finite-dimensional parameter vector $\theta_{t}$. By construction, $\sum_{j=0}^{J}q_{j}(x,\theta_{j})=1$. For example, when treatments are multi-valued, qualitatively distinct, and without a logical ordering, a popular specification for $q_{t}\left(  x,\theta_{t}\right)$ is the multinomial logit model\footnote{One limitation with the multinomial logit model is the  independence of irrelevant alternatives (IIA) assumption imposed by it. One alternative to avoid the IIA assumption is the multinomial probit model. But if $J\geq 2$, multinomial probit model meets practical obstacles as it involves computing $(J+1)$-dimensional integrals.} (e.g., \citealp{Imbens2000}),
\begin{equation}
q_{t}\left(  X,\theta_{t}\right)  =\frac{\exp\left(  X^{\prime}\theta_{t}\right)  }{1+\sum_{j=1}^{J}\exp\left(  X^{\prime}\theta_{j}\right)}\text{, }\,\,\,t\in\left\{  1,\dots,J\right\}  .\label{eq:mult_logit}
\end{equation}

The parameter vector $\left\{  \theta_{t},t\in\mathcal{T}\right\}  $ can be estimated using the maximum likelihood approach. Researchers can then estimate $\beta(t,s)$ through the inverse probability weighting (IPW) approach.

Given the widespread empirical practice of adopting parametric models for the GPS, a natural concern is potential model misspecifications. In particular, if the working model $q_{t}\left(x,\theta_{t}\right)$ for the GPS $p_{t}\left(x\right)$ is misspecified, causal effects estimators such as the inverse probability weighted estimator are in general biased and policy recommendations based on them can be highly misleading (e.g., \citealp{Linden2016}). To assess whether this is the case, one can frame this as a specification test  for $q_{t}\left(x,\theta_{t}\right)$ that fits into our testing framework by simply setting $e(t;\theta)$ to be the parametrically specified generalized error under the null for every $t\in\mathcal T$, i.e., $e(t;\theta)\equiv1(T=t)-q_{t}(X,\theta)$.

For motivational and concreteness sake, we take $e(t;\theta)$ as the generalized error of the propensity score model as mentioned above in the rest of the paper, though we can handle much more general models. See, e.g., Remark \ref{rem:general} for additional details.

\subsection{Our proposed test}\label{sec:dim_red}

The characterization of $H_{0}$ and $H_{1}$ in (\ref{cmr}) and (\ref{cmr_alter}), respectively, makes explicit that we are interested in testing conditional moment restrictions (\citealp{Gonzalez-Manteiga2013}). As argued by \cite{Escanciano2006b},
(\ref{cmr}) can be equivalently characterized as
\begin{align}
H_{0}:R_{t}^{pro}(\beta,u;\theta_0)=0&\text{ almost everywhere (a.e.) }\left(\beta^{\top},u\right)  ^{\top}\in\Pi_{pro},\notag\\
&\text{for some }\,\theta_0\in\Theta\subset\mathbb{R}^{d_\theta}\text{ and all }t\in\mathcal{T}, \label{null}
\end{align}
where
\[
R_{t}^{pro}(\beta,u;\theta)\equiv\mathbb{E}\left[  e(t;\theta)1\left(X^\top\beta\leq u\right)  \right]  ,
\]
and $\Pi_{pro}\equiv\mathbb{S}^{d_{x}}\times\overline{\mathbb R}$ denotes the projected space, with $\overline{\mathbb R}=\lbrack-\infty,\infty]$ the extended real line and $\mathbb{S}^{d_{x}}$ the unit ball in $\mathbb{R}^{d_{x}}$, i.e., $\mathbb{S}^{d_{x}}=\left\{  \beta\in\mathbb{R}^{d_{x}}:||\beta||=1\right\}  $ with $\left\vert \left\vert A\right\vert \right\vert =\left[  \text{tr}\left(  AA^{\top}\right)  \right]
^{1/2}$ denoting the Euclidean norm for a generic matrix $A$.

Although one can find alternative characterizations of $H_{0}$, see, e.g., \cite{Bierens1982}, \cite{Stute1997}, and \cite{Dominguez2004}, our main motivations for expressing $H_{0}$ as in (\ref{null}) are that $\left(  i\right)  $ $R_{t}^{pro}(\beta,u;\theta_0)$ is based on unconditional moment restrictions, implying that we can avoid the use of tuning parameters such as bandwidths when estimating $R_{t}^{pro}(\beta,u;\theta_0)$; and $\left(  ii\right)  ~R_{t}^{pro}(\beta,u;\theta_0)$ depends on covariates only through the one-dimensional projection $X^\top\beta$, greatly reducing the dimensionality of the problem. Indeed, this dimension-reduction device has been proven valuable in various contexts that need to deal with many covariates; see, e.g., \cite{Escanciano2006b}, \cite{Garcia-Portugues2014}, \cite{Sun2017}, \cite{Zhu2017}, and \cite{Kim2020}; for an overview, see \cite{Guo2017}. However, it is worth mentioning that (\ref{null}) involves not only a single process $R_{t}^{pro}(\beta,u;\theta_0)$ as is commonly the case in the
specification testing literature (see \citealp{Escanciano2008} for an exception), but $J$ different processes $R_{t}^{pro}(\beta,u;\theta_0)$ associated with the $J$ different treatment levels $t$.

From (\ref{null}), one natural way to proceed is to compute the generalized residual marked empirical process based on the projections $1\left(X^\top\beta\leq u\right)$,
\[
R_{n,t}^{pro}(\beta,u;\widehat{\theta}_{n})=\frac{1}{\sqrt{n}}\sum_{i=1}^{n}e_{i}(t;\widehat{\theta}_{n})1(X_i^\top\beta\leq u),\,\,\,\left(\beta^{\top},u\right)  ^{\top}\in\Pi_{pro},
\]
where $\widehat{\theta}_{n}$ is any $\sqrt{n}$-consistent estimator for $\theta_0$, say the maximum likelihood estimator, and $e_{i}(t;\widehat{\theta}_{n})\equiv1(T_{i}=t)-q_{t}(X_{i},\widehat{\theta}_{n})$, $i=1,\ldots,n$, are the parametrically specified generalized residuals under the null $H_{0}$. Then, one can use continuous functionals of $R_{n,t}^{pro}(\beta,u;\widehat{\theta}_{n})$ to measure its distance from zero and assess if $H_{0}$ is rejected or not. 

Although natural, there are some potential drawbacks of using $R_{n,t}^{pro}(\beta,u;\widehat{\theta}_{n})$. For instance, the underlying null limiting distribution of $R_{n,t}^{pro}(\beta,u;\widehat{\theta}_{n})$ depends on the estimator $\widehat{\theta}_{n}$. Indeed, the asymptotic null distribution of tests based on $R_{n,t}^{pro}(\beta,u;\widehat{\theta}_{n})$ will depend on whether one estimates $\theta_0$ using maximum likelihood, nonlinear least squares, or by the method of estimating equations, even though the underlying specification for the null model is the same across these estimation methods. Perhaps more importantly, as noted by \cite{Escanciano2006b}, tests based on $R_{n,t}^{pro}(\beta,u;\widehat{\theta}_{n})$ also require that $\sqrt{n}(\widehat{\theta}_{n}-\theta_0)$ admits an asymptotically linear representation.\footnote{Throughout the paper, an estimator $\widehat{\theta}_{n}$ is said to be root-$n$ asymptotically linear if it satisfies the following asymptotic expansion under $H_0$:
\begin{equation*}
\sqrt n\left(\widehat\theta_n-\theta_0\right)=\frac{1}{\sqrt n}\sum_{i=1}^nl(T_i,X_i,\theta_0)+o_p(1),
\end{equation*}
where $l(\cdot,\cdot,\cdot)$ is such that $\mathbb E[l(T,X,\theta_0)]=0$ and $L(\theta_0)=\mathbb E[l(T,X,\theta_0)l(T,X,\theta_0)^\top]$ exists and is positive definite. Note that different $\widehat\theta_n$ may have different $l(\cdot,\cdot,\cdot)$.} Such a condition can be demanding, especially when one wishes to use estimation methods that involve penalization (see, e.g., \citealp{Knight2000b} and \citealp{Buhlmann2011}). The fact that $R_{n,t}^{pro}(\beta,u;\widehat{\theta}_{n})$ is not invariant to $\widehat{\theta}_{n}$ also precludes an easy-to-implement multiplier bootstrap procedure to obtain critical values. This is inconvenient, especially when the response variables are discrete or mixed, since the wild bootstrap method requires regenerating dependent variables and fails to mimic the original data structure.

Given these potential drawbacks, we follow an alternative route that leads to estimator-invariant tests. More
specifically, our proposed test statistics are continuous functionals of the following generalized residual marked empirical process based on the double projections:
\[
R_{n,t}^{dpro}(\beta,u;\widehat{\theta}_{n})\equiv\frac{1}{\sqrt{n}}\sum_{i=1}^{n}e_{i}(t;\widehat{\theta}_{n})\mathcal{P}_{n,t}1(X_{i}^\top\beta\leq u),\,\,\,\left(  \beta^{\top},u\right)  ^{\top}\in\Pi_{pro},
\]
where the double-projected weight is
\begin{equation}
\mathcal{P}_{n,t}1\left(X_i^\top\beta\leq u\right)  \equiv1\left(X_i^\top\beta\leq u\right)  -g_{t}(X_i,\widehat{\theta}_{n})^\top\Delta_{n,t}^{-1}(\widehat{\theta}_{n})G_{n,t}(\beta,u;\widehat{\theta}_{n}),
\label{dp_weight}%
\end{equation}
where, for each $t\in\mathcal{T}$, $g_{t}(x,\theta)\equiv\partial q_{t}(x,\theta)/\partial\theta$, denotes the score function associated with the parametric model $q_{t}(x,\theta)$, and
\[
\Delta_{n,t}(\widehat{\theta}_{n})=\frac{1}{n}\sum_{i=1}^{n}g_{t}(X_{i},\widehat{\theta}_{n})g_{t}^{\top}(X_{i},\widehat{\theta}_{n})\text{ and } G_{n,t}(\beta,u;\widehat{\theta}_{n})=\frac{1}{n}\sum_{i=1}^{n}g_{t}(X_{i},\widehat{\theta}_{n})1\left(X_{i}^\top\beta\leq u\right)  .
\]

We label $\mathcal{P}_{n,t}1\left(X^\top\beta\leq u\right)  $ as a double-projection because, as it is evident from (\ref{dp_weight}), it involves first using the projection proposed by \cite{Escanciano2006b}, $1\left(X^\top\beta\leq u\right)  $, and then projecting $1\left(X^\top\beta\leq u\right)  $ onto the tangent space of the nuisance parameters (see, e.g., \citealp{Neyman1959}, \citealp{Escanciano2014a}, and \citealp{SantAnna2019}). To the best of our knowledge, this paper is the first to incorporate this double-projection argument, which, in practice, translates to test statistics that are robust against the ``curse of dimensionality'' and whose limiting null distributions are asymptotically invariant to $\widehat{\theta}_{n}$. This latter property follows from the fact that, for each $t\in\mathcal{T}$,
\[
\mathbb{E}\left[  g_{t}(X,\theta_0)\mathcal{P}_{t}1\left(X^\top\beta\leq u\right)  \right]  \equiv0,
\]
almost everywhere in $\left(  \beta^{\top},u\right)  ^{\top}\in\Pi_{pro}$, where
\begin{equation}
\mathcal{P}_{t}1\left(X^\top\beta\leq u\right)  \equiv1\left(X^\top\beta\leq u\right)  -g_{t}(X,\theta_0)^\top\Delta_{t}^{-1}(\theta_0)G_{t}(\beta,u;\theta_0), \label{projection}%
\end{equation}
with $\Delta_{t}(\theta)=\mathbb{E}\left[  g_{t}(X,\theta)g_{t}(X,\theta)^\top\right]  $ and $G_{t}(\beta,u;\theta)=\mathbb{E}\left[
g_{t}(X,\theta)1\left(X^\top\beta\leq u\right)  \right]  $.  Note also that $R_{n,t}^{dpro}(\beta,u;\widehat{\theta}_{n})$ does not depend on tuning parameters such as bandwidths. 

The intuition behind (\ref{projection}) is very simple. First of all, note that, for each $t\in\mathcal{T}$, $\left(  \beta^{\top},u\right)  ^{\top}\in\Pi_{pro}$, $\Delta_{t}^{-1}\left(  \theta_0\right)  G_{t}\left(\beta,u;\theta_0\right)  $ is the vector of linear projection coefficients of regressing $1\left(X^\top\beta\leq u\right)  $ on the score function $g_{t}(X,\theta_0)$. Thus, it follows that
\[
g_{t}(X,\theta_0)^\top\Delta_{t}^{-1}\left(  \theta_0\right)G_{t}\left(  \beta,u;\theta_0\right)
\]
is the best linear predictor of $1\left(X^\top\beta\leq u\right)  $ given $g_{t}(X,\theta_0)$, and that (\ref{projection}) is nothing more than the associated projection error, which, by definition, is orthogonal to
$g_{t}(X,\theta_0)$. 

This orthogonality condition exploited by the double-projection procedure has important consequences. For example, under some weak regularity conditions, uniformly in $\left(  \beta^{\top},u\right)  ^{\top}\in\Pi_{pro}$,
\begin{equation}
R_{n,t}^{dpro}(\beta,u;\widehat{\theta}_{n})=\frac{1}{\sqrt{n}}\sum_{i=1}^{n}e_{i}(t;\theta_0)\mathcal{P}_{t}1(X_{i}^\top\beta\leq u)+o_{p}(1), \label{equivalence2_dpro}
\end{equation}
for each $t\in\mathcal{T}$; see the proof of Theorem \ref{theorem1} in the next section. As so, $R_{n,t}^{dpro}(\beta,u;\widehat{\theta}_{n})$ is asymptotically invariant to the choice of the estimator $\widehat{\theta}_{n}$, which, in turn, facilitates a simple multiplier bootstrap method to simulate critical values of test statistics based on $R_{n,t}^{dpro}(\beta,u;\widehat{\theta}_{n})$. Indeed, given that we can ``ignore'' estimation effects when computing $R_{n,t}^{dpro}(\beta,u;\widehat{\theta}_{n})$, all we need to do is ``perturbate'' the residuals $e(t;\widehat{\theta}_{n})$; see Section \ref{boot} for details. Here, it is worth stressing that this is only feasible due to the usage of the second projection. Without it, we would need to either rule out discrete/mixed outcomes, and/or further impose that $\sqrt n(\widehat\theta_n-\theta_0)$ admits an asymptotically linear representation. Even in these cases, different estimators usually have different asymptotically linear representations, leading to different multiplier bootstraps with potentially different testing results. Of course, (\ref{equivalence2_dpro}) allows us to avoid these problems.

In order to operationalize our testing procedure, we need to choose a norm to measure the distance of $R_{n,t}^{dpro}(\beta,u;\widehat\theta_n)$ from zero. We propose using the popular Cram\'{e}r--von Mises ($CvM$ thereafter)-type test statistic
\begin{equation}
CvM_{n}^{dpro}=\sum_{t\in\mathcal{T}}a_{n}(t)\int_{\Pi_{pro}}\left(R_{n,t}^{dpro}(\beta,u;\widehat{\theta}_{n})\right)  ^{2}\,F_{n,\beta}(du)\,d\beta, \label{cvm}%
\end{equation}
where, for each $t$, $a_{n}(t)$ is a pre-specified (potentially random) non-negative weighting function, $F_{n,\beta}(u)=n^{-1}\sum_{i=1}^{n}1\left(X_{i}^\top\beta\leq u\right)$ is the empirical distribution function of the one-dimensional projected regressors $\left\{X^\top_{i}\beta\right\}_{i=1}^{n}$ for any fixed projected direction $\beta\in\mathbb{S}^{d_{x}}$, and $d\beta$ is the rescaled uniform density on the unit sphere $\mathbb{S}^{d_{x}}$. 

We reject the null $H_0$ in favor of the alternative $H_1$ whenever $CvM_{n}^{dpro}$ in \eqref{cvm} is ``overly'' large. For the sake of practical convenience, in the empirical application in Section \ref{empirical_section} and Monte Carlo simulations in the online supplementary appendix, we use the constant weight $a_{n}(t)\equiv1$ for all $t\in\mathcal{T}$, though other sensible data-driven choices are feasible, e.g., $a_{n}(t)=n^{-1}\sum_{i=1}^{n}1(T_{i}=t)$. Due to space constraint, we do not intend to search for the optimal weight that yields the highest power.

At this stage, one may wonder why we have chosen to use a $CvM$-type instead of a Kolmogorov--Sminov-type test statistic. The reason is computational: as we show below in Lemma \ref{lemma1}, (\ref{cvm}) can be written in a closed-form expression and does not rely on any numerical integration method. Furthermore, one does not need to compute a different projection for each $\left(  \beta^{\top},u\right)  ^{\top}\in\Pi_{pro}$, which is arguably a ``natural'' step if one were to take \citet{Neyman1959}'s ``debiasing'' proposal literally. A direct consequence of these attractive computational features is that (\ref{cvm}) can be easily implemented even with many covariates and many treatment levels. In addition, the closed-form expression given in Lemma \ref{lemma1} can readily be used to calculate the multiplier bootstrapped version of our $CvM$ test statistic in Section \ref{boot}.

\begin{lemma}
\label{lemma1} Let $CvM_{n}^{dpro}$ be defined in \eqref{cvm} with $\mathbb{S}^{d_{x}}$ the $d_{x}$-dimensional unit sphere. Then, we have
\begin{equation}
CvM_{n}^{dpro}=\sum_{t\in\mathcal{T}}a_{n}(t)\frac{1}{n^{2}}\sum_{i=1}^{n}\sum_{j=1}^{n}\sum_{r=1}^{n}e_{i}^{pro}(t;\widehat{\theta}_{n})e_{j}%
^{pro}(t;\widehat{\theta}_{n})A_{ijr}, \label{cvm_closed}
\end{equation}
with
\begin{equation}
e_{i}^{pro}(t;\widehat{\theta}_{n})=e_{i}(t;\widehat{\theta}_{n})-g_{t}X_{i},\widehat{\theta}_{n})^\top\Delta_{n,t}^{-1}(\widehat{\theta}_{n})\frac{1}{n}\sum_{s=1}^{n}g_{t}(X_{s},\widehat{\theta}_{n})e_{s}(t;\widehat{\theta}_{n}), \quad i=1,\ldots,n,\label{e_proj}%
\end{equation}
and
\[
A_{ijr}=\int_{\mathbb{S}^{d_{x}}}1\left(X_{i}^\top\beta\leq X_{r}^\top\beta\right)  1\left(X^\top_{j}\beta\leq X^\top_{r}\beta\right)d\beta=A_{ijr}^{\left(  0\right)  }\frac{\pi^{d_{x}/2-1}}{\Gamma\left(d_{x}/2\right)  },
\]
where $\Gamma\left(  \cdot\right)  $ is the gamma function, $\arccos$ is theinverse cosine function, and
\[
A_{ijr}^{\left(  0\right)  }=\left\{
\begin{array}
[c]{ll}%
2\pi, & if~X_{i}=X_{r}=X_{j},\\
\pi, & if\text{ }X_{i}=X_{j},~or~X_{i}=X_{r},~or~X_{j}=X_{r},\\
\left|\pi-\arccos\left(  \dfrac{\left(  X_{i}-X_{r}\right)  ^\top\left(
X_{j}-X_{r}\right)  }{\left\Vert X_{i}-X_{r}\right\Vert \left\Vert X_{j}%
-X_{r}\right\Vert }\right)\right|,  & otherwise.
\end{array}
\right.
\]
\end{lemma}

It is interesting to note that $e_{i}^{pro}(t;\widehat{\theta}_{n})$ for $i=1,\ldots,n$ in \eqref{e_proj} are simply the ordinary least squares residuals from regressing $e_{i}(t;\widehat{\theta}_{n})$ on $g_{t}(X_{i},\widehat{\theta}_{n})$. Lemma \ref{lemma1} yields an explicit formula for our double-projected test statistic $CvM_{n}^{dpro}$. It builds on \cite{Escanciano2006b} and \cite{Garcia-Portugues2014}, who derived expressions for the $CvM$-type functionals of ``single-projected'' empirical processes akin to $R_{n,t}^{pro}(\beta,u;\widehat{\theta}_{n})$. In fact, it can be shown that the  $CvM$ test statistic based on $R_{n,t}^{pro}(\beta,u;\widehat{\theta}_{n})$ has the following closed-form expression:
\begin{align*}
CvM_n^{pro}=&\sum_{t\in\mathcal{T}}a_{n}(t)\int_{\Pi_{pro}}\left(R^{pro}_{n,t}(\beta,u;\widehat{\theta}_{n})\right)^{2}\,F_{n,\beta}(du)\,d\beta\\
=&\sum_{t\in\mathcal{T}}a_{n}(t)\frac{1}{n^{2}}\sum_{i=1}^{n}\sum_{j=1}^{n}\sum_{r=1}^{n}e_{i}(t;\widehat{\theta}_{n})e_{j}(t;\widehat{\theta}_{n})A_{ijr},
\end{align*}
which is robust to the dimensionality $d_x$ of $X$ but not robust to the choice of estimator $\widehat\theta_n$. With the introduction of the second projection, we are able to obtain a $CvM$ test statistic that is robust to both dimensionality $d_x$ and estimator $\widehat\theta_n$.

Just like in \cite{Escanciano2006b} and \cite{Garcia-Portugues2014}, $A_{ijr}$ represents the surface area of particular spherical regions depending on whether $X$'s are the same across observations: it is the whole sphere $\mathbb{S}^{d_{x}}$ when $X_i=X_r=X_j$, a hemisphere of $\mathbb{S}^{d_{x}}$ when $X_i=X_j$ or $X_i=X_r$ or $X_i=X_r$ (but not all $X$'s are the same), or a spherical wedge of width angle given by the third entry of $A_{ijr}^{(0)}$. Thus, as discussed in \cite{Escanciano2006b} and \cite{Garcia-Portugues2014}, we can express $A_{ijr} = A_{ijr}^{\left(  0\right)  }\left.{\pi^{d_{x}/2-1}}\right/{\Gamma\left(d_{x}/2\right)  }$, as in Lemma \ref{lemma1} above. 

In connection with the expression of $CvM_n^{pro}$, what is new in Lemma \ref{lemma1} is that, instead of computing infinitely many projected residuals, one for each $\left(  \beta^{\top},u\right)  ^{\top}\in\Pi_{pro}$ as it is suggested by (\ref{dp_weight}), it suffices to use the sequence of projected\ parametric residuals $\left\{e_{i}^{pro}(t;\widehat{\theta}_{n}),\,i=1,\dots n,\,t\in\mathcal{T}\right\}$ as defined in (\ref{e_proj}), which do not depend on the projection direction $\beta$. Note that the construction of $e_{i}^{pro}(t;\widehat{\theta}_{n})$ and its intuitive interpretation are only possible with the help of the second projection. 

\section{Asymptotic results\label{asy}}

\subsection{Asymptotic null distribution\label{asy_null}}

In this section, we formally investigate the limiting behavior of double-projected generalized residual marked empirical process $R_{n,t}^{dpro}(\beta,u;\widehat{\theta }_{n})$ under the null hypothesis $H_{0}$ in (\ref{cmr}) and consequently that of the test statistic $CvM_{n}^{dpro}$ based on it.

First, let us denote by $F_{X}(\cdot)$ the cumulative distribution function (CDF) of covariates $X$. Also, let $\Psi_{pro}(du,d\beta)=F_{\beta}(du)d\beta$. Recall that $p_{t}(x)=\mathbb{P}(T=t|X=x)$ for every $t\in\mathcal{T}$ are the true but unknown GPS.\footnote{One can alternatively denote $p_{t}(x)$ as the true but unknown model underlying a conditional moment restriction; see Remark \ref{rem:general}.}  We list all the relevant regularity conditions as follows.

\begin{assumption}
\label{A1}The random sample $\left\{  \left(  X_{i}^{\top},T_{i}\right)  ^{\top},i=1\dots n\right\}  $ consists of a sequence of independent and identically distributed random vectors from $(X^{\top},T)^{\top}$. 
\end{assumption}

\begin{assumption}
\label{A2} For each $t \in \mathcal{T}\subseteq \mathbb{N}$, the propensity score model $q_{t}(X,\theta)$ is known up to the finite-dimensional parameter $\theta$, and is twice continuously differentiable in a neighborhood $\Theta_{0}$ of $\theta_0$ with $\Theta_{0}\subset\Theta$. The score function $g_{t}(X,\theta)=\partial q_{t}(X,\theta)/\partial\theta$ satisfies that there exists a $F_{X}(\cdot)$-integrable function $M(\cdot)$ such that $\sup_{\theta\in\Theta_{0}}\left\vert \left\vert g_{t}(\cdot,\theta)\right\vert \right\vert \leq M(\cdot)$.
\end{assumption}

\begin{assumption}
\label{A3}(i) The parameter space $\Theta$ is a compact subset of $\mathbb{R}^{d_\theta}$; (ii) the true parameter $\theta_0$ belongs to the interior of $\Theta$; and (iii) the estimator $\widehat{\theta}_{n}$ satisfies that $\left\vert \left\vert \widehat{\theta}_{n}-\theta_0\right\vert\right\vert =O_{p}(n^{-1/2})$ under $H_0$, and $\left\vert \left\vert \widehat{\theta}_{n}-\theta^{\ast}\right\vert\right\vert =o_{p}(1)$ under $H_1$ for some $\theta^\ast$ in the parameter space $\Theta$.
\end{assumption}

\begin{assumption}
\label{A4}The integrating function $\Psi_{pro}(\cdot)$ is absolutely continuous with respect to the Lebesgue measure on $\Pi_{pro}$.
\end{assumption}

Assumptions \ref{A1}-\ref{A4} are standard in the specification testing literature; see, e.g., \cite{Escanciano2006b}. Note that we only require the root-$n$ consistency of $\widehat\theta_n$ in Assumption \ref{A3}(iii) rather than $\widehat\theta_n$ being root-$n$ asymptotically linear as required by \citet[Assumption A3(b)]{Escanciano2006b}.

To present our asymptotic results, henceforth, we adopt the following notation. For a generic set $\mathcal{G}$, let $l^{\infty}\left( \mathcal{G}\right)  $ be the Banach space of all uniformly bounded real
functions on $\mathcal{G}$, equipped with the uniform metric $\left\Vert f\right\Vert _{\mathcal{G}}\equiv\sup_{z\in\mathcal{G}}\left\vert f\left( z\right)  \right\vert $. We study the weak convergence of $R_{n,t}^{dpro}(\beta,u;\widehat{\theta}_{n})$ and its related processes as elements of $l^{\infty}\left(  \Pi_{pro}\right)  $, where $\Pi_{pro}\equiv\mathbb{S}^{d_{x}}\times\left[  -\infty,\infty\right]  $ with $\mathbb{S}^{d_{x}}$ the unit ball in $\mathbb{R}^{d_{x}}$. Let ``$\Longrightarrow$'' denote weak convergence on $\left( l^{\infty}\left(  \Pi_{pro}\right)  ,\mathcal{B}_{\infty}\right)  $ in the sense of J. Hoffmann--J$\phi$rgensen, where $\mathcal{B}_{\infty}$ denotes the corresponding Borel $\sigma$-algebra - see e.g. Definition 1.3.3 in \cite{VanderVaart1996}. Then it is shown that under the null $R_{n,t}^{dpro}(\cdot,\cdot;\widehat{\theta}_{n})\Longrightarrow R_{\infty,t}^{dpro}$, a centered Gaussian process with its covariance structure defined in Theorem \ref{theorem1}. 

The true generalized error is defined as $\varepsilon(t)=1(T=t)-p_{t}(X)$, which satisfies $\mathbb E[\varepsilon(t)|X]=0$ almost surely ($a.s.$) for each $t\in\mathcal T$ regardless of whether the null hypothesis is true. We state formally the asymptotic null behavior of our test statistic $CvM_{n}^{dpro}$ in the following theorem.

\begin{theorem}
\label{theorem1} Suppose Assumptions \ref{A1}-\ref{A4} hold. Then, under the null hypothesis $H_{0}$ in (\ref{cmr}), for any sequence $a_{n}(t)=a(t)+o_{p}\left(  1\right)  $, with $a(t)>0$ and $0<\sum_{t\in\mathcal{T}}a(t)\leq C<\infty$, we have that
\[
CvM_{n}^{dpro}\xrightarrow{d}CvM_{\infty}^{dpro}\equiv\sum_{t\in\mathcal{T}}a(t)\int_{\Pi_{pro}}\left(  R_{\infty,t}^{dpro}\left(  \beta,u\right)\right)  ^{2}\,\Psi_{pro}\left(  du,d\beta\right)  ,
\]
where $R_{\infty,t}^{dpro}$ is a Gaussian process with mean zero and covariance structure
\begin{equation}
\mathbb{K}_{t}^{dpro}\left(  \left(  \beta,u\right)  ,\left(  \beta^{\prime},u^{\prime}\right)  \right)  =\mathbb{E}\left[  \sigma_{t}^{2}(X)\mathcal{P}_{t}1\left(X^\top\beta\leq u\right)  \mathcal{P}_{t}1\left(X^\top\beta'\leq u^{\prime}\right)  \right]  , \label{kw}
\end{equation}
where $\sigma_{t}^{2}(X)=\mathbb{E}\left[  \varepsilon^{2}(t)|X\right]=p_{t}(X)\left(  1-p_{t}(X)\right)  =q_{t}(X,\theta_0)\left(1-q_{t}(X,\theta_0)\right)  $ is the conditional variance function of generalized error $\varepsilon(t)$ given $X$ under the null.
\end{theorem}

To prove Theorem \ref{theorem1}, we first show that the asymptotic null behavior of $R_{n,t}^{dpro}(\beta,u;\widehat{\theta}_{n})$ does not depend on $\widehat{\theta}_{n}$. Based on this result, we combine the weak convergence of the doubly-projected empirical process $R_{n,t}^{dpro}$ with the continuous mapping theorem (see, e.g., \citealp{VanderVaart1996}, Theorem 1.3.6) to derive the asymptotic distribution of our proposed Cram\'{e}r--von Mises test statistic $CvM_{n}^{dpro}$ under the null $H_0$.

\subsection{Asymptotic power\label{asy_alter}}

In this section, we study the asymptotic power properties of the $CvM_{n}^{dpro}$ test statistic based on $R_{n,t}^{dpro}(\beta,u;\widehat{\theta}_{n})$ under the fixed (i.e., global) alternative and a certain sequence of local alternatives converging to the null $H_0$ at the usual parametric rate. We first consider the fixed alternative hypothesis $H_{1}$ in (\ref{cmr_alter}).

\begin{theorem}
\label{thh1} Suppose Assumptions \ref{A1}-\ref{A4} hold. Then, under the fixed alternative hypothesis $H_{1}$ in (\ref{cmr_alter}), for any sequence $a_{n}(t)=a(t)+o_{p}\left(  1\right)  $, with $a(t)>0$ and $0<\sum_{t\in\mathcal{T}}a(t)\leq C<\infty$, we have that
\[
\frac{CvM_{n}^{dpro}}{n}\xrightarrow{p}\sum_{t\in\mathcal{T}}a(t)\int_{\Pi_{pro}}\left(  \mathbb{E}\left[  \left(  p_{t}\left(  X\right)-q_{t}\left(  X,\theta^{\ast}\right)  \right)  \mathcal{P}_{t}1\left(X^\top\beta\leq u\right)  \right]  \right)  ^{2}\,\Psi_{pro}\left(du,d\beta\right)  .
\]

\end{theorem}

It follows from Theorem \ref{thh1} that, under the fixed alternative alternative $H_{1}$, as long as the unconditional expectation
\[
\mathbb{E}\left[  \left(  p_{t}\left(  X\right)  -q_{t}\left(  X,\theta^{\ast}\right)  \right)  \mathcal{P}_{t}1\left(X^\top\beta\leq u\right)\right]  \neq0
\]
for some $(\beta^\top,u)^\top$ and for some treatment level $t\in\mathcal{T}$, $CvM_{n}^{dpro}$ will diverge to positive infinity at the $n$ rate, indicating that $CvM_{n}^{dpro}$ is able to detect such a fixed alternative with probability tending to one. On the other hand, $CvM_{n}^{dpro}$ might not be consistent against all fixed alternative hypotheses in (\ref{cmr_alter}) if $\mathbb{E}\left[  \left(  p_{t}\left(  X\right)  -q_{t}\left(X,\theta^{\ast}\right)  \right)  \mathcal{P}_{t}1\left(X^\top\beta\leq u\right)  \right]  =0$ for every $t\in\mathcal T$. Specifically, the test statistic cannot distinguish those alternatives such that, for every $t\in\mathcal{T}$, the difference between $p_{t}(X)$ and $q_{t}(X,\theta^{\ast})$ is collinear to the score function $g_{t}(X,\theta^{\ast})$ associated with $q_{t}(X,\theta^{\ast})$. However, we do not think this type of alternative is of main empirical concern. If one were concerned with this, one could use alternative testing procedures such as \citet{Stute1997}, \cite{Escanciano2006b}, and \citet{dominguez2020specification}.

We now proceed to consider the asymptotic local power properties of our proposed test statistic. Toward this end, we study the asymptotic distribution of $R_{n,t}^{dpro}(\beta,u;\widehat{\theta}_{n})$ under a certain sequence of Pitman-type local alternatives converging to the null at a parametric rate. In particular, we consider the data-generating process for the sequence of local alternatives given by
\begin{equation}
H_{1,n}:\mathbb{P}\left[  p_{t}(X)=q_{t}(X,\theta_0)+\frac{r_{t}(X)}{\sqrt{n}}\right]  =1\text{ for some }\theta_0\in\Theta\in\mathbb{R}^{d_\theta}\text{ and all }t\in\mathcal{T}, \label{h1n}
\end{equation}
where, for each $t\in\mathcal{T},$ the direction of departure from $H_{0}$ is given by function $r_{t}(X)$ (potentially different for each $t$), which is assumed to be $F_{X}(\cdot)$-integrable with zero mean and satisfy $\mathbb{P}\left(  r_{t}(X)=0\right)  <1$. 

\begin{theorem}
\label{thh1n} Suppose Assumptions \ref{A1}-\ref{A4} hold. Then, under the sequence of local alternatives $H_{1,n}$ in (\ref{h1n}), for any sequence $a_{n}(t)=a(t)+o_{p}\left(  1\right)  $, with $a(t)>0$ and $0<\sum_{t\in\mathcal{T}}a(t)\leq C<\infty$, we have that
\[
CvM_{n}^{dpro}\xrightarrow{d}CvM_{1,\infty}^{dpro}\equiv\sum_{t\in\mathcal{T}}a(t)\int_{\Pi_{pro}}\left(  R_{\infty,t}^{dpro}\left(  \beta,u\right)+\delta_{t}\left(  \beta,u\right)  \right)  ^{2}\,\Psi_{pro}\left(
du,d\beta\right)  ,
\]
where $R_{\infty,t}^{dpro}$ is the same Gaussian process as defined in Theorem \ref{theorem1}, and $\delta_{t}$ is a deterministic shift function given by
\[
\delta_{t}(\beta,u)=\mathbb{E}\left[  r_{t}\left(  X\right)  \mathcal{P}_{t}1\left(X^\top\beta\leq u\right)  \right]  .
\]

\end{theorem}

An immediate consequence of Theorem \ref{thh1n} is that whenever there exists some $t\in\mathcal{T}$ such that the deterministic shift function $\delta_{t}\left(  \beta,u\right)  \not =0$ for at least some $(\beta^\top,u)^\top\in\Pi_{pro}$ with a positive Lebesgue measure, our proposed Cram\'{e}r--von Mises test
statistic will have non-trivial power against local alternatives of the form (\ref{h1n}). A pathological situation in which our test will only have trivial local power against such local alternatives is when $r_{t}(X)$ is a linear combination of score function $g_{t}(X,\theta_0)$ for every treatment level $t\in\mathcal{T}$, i.e., $r_{t}(X)=\nu^{\top}g_{t}(X,\theta_0)$ $a.s.$ for some nonzero vector $\nu$. In such a case, the limiting distribution of $CvM_{n}^{dpro}$ under $H_{0}$ and $H_{1,n}$ is the same so that $H_{1,n}$ cannot be detected. However, such a specific class of local alternatives is arguably of limited practical interest. 

\section{A multiplier bootstrap procedure\label{boot}}

Since the limiting null distribution of our test statistic $CvM_{n}^{dpro}$ is non-pivotal, we propose a simple-to-implement multiplier bootstrap procedure to obtain the bootstrapped $p$-values or critical values and show its asymptotic validity. Below is its implementation:

\begin{enumerate}
\item For each $t\in\mathcal{T}$, $i=1,\dots n$, generate $e_{i}^{\ast}(t;\widehat{\theta}_{n})=V_{i}\,e_{i}(t;\widehat{\theta}_{n}),$ where $\{V_{i},i=1,\dots,n\}$ is a sequence of independent and identically distributed random variables with mean zero, variance one, and finite third moment; e.g., Rademacher random variables with $\mathbb{P}\left(  V=-1\right)=\mathbb{P}\left(V=1\right)  =1/2$ (\citealp{Liu1988}) or Bernoulli random variable with $\mathbb{P}\left(  V=1-\kappa\right)  =\kappa/\sqrt{5}$ and $\mathbb{P}\left(V=\kappa\right)  =1-\kappa/\sqrt{5}$, where $\kappa=\left(  \sqrt 5+1\right)/2$ (\citealp{Mammen1993}).

\item Compute
\begin{align}
\left(  CvM_{n}^{dpro,\ast}\right)  ^{b}=\sum_{t\in\mathcal{T}}a_{n}(t)\frac{1}{n^{2}}\sum_{i=1}^{n}\sum_{j=1}^{n}\sum_{r=1}^{n}e_{i}^{pro,\ast}(t;\widehat{\theta}_{n})e_{j}^{pro,\ast}(t;\widehat{\theta}_{n})A_{ijr}, \label{cvm_closed_boot}
\end{align}
where
\begin{equation}
e_{i}^{pro,\ast}(t;\widehat{\theta}_{n})=e_{i}^{\ast}(t;\widehat{\theta}_{n})-g_{t}^{\top}(X_{i},\widehat{\theta}_{n})\Delta_{n,t}^{-1}(\widehat{\theta}_{n})\frac{1}{n}\sum_{s=1}^{n}g_{t}(X_{s},\widehat{\theta}_{n})e_{s}^{\ast}(t;\widehat{\theta}_{n}),\quad i=1,\ldots,n. \label{residual_projection_boot}
\end{equation}

\item Repeat Steps 1 and 2 $B$ times, and collect $\left\{  \left(CvM_{n}^{dpro,\ast}\right)  ^{b},b=1\dots,B\right\}  .$

\item Obtain the $\left(  1-\alpha\right)  $ quantile of $\left\{  \left(CvM_{n}^{dpro,\ast}\right)  ^{b},b=1\dots,B\right\}$, $c_{n,\alpha}^{\ast}$, and set it as the critical value for the test with significance level $\alpha $, for $0<\alpha<1$.

\item Reject the null hypothesis $H_0$ in (\ref{cmr}) if $CvM_{n}^{dpro}$ is greater than the critical value $c_{n,\alpha}^{\ast}$, and fail to reject (\ref{cmr}) otherwise.
\end{enumerate}

Note that $e_{i}^{pro,\ast}(t;\widehat{\theta}_{n})$ for $i=1,\ldots,n$ in \eqref{residual_projection_boot} are simply the ordinary least squares residuals from regressing $e_{i}^{\ast}(t;\widehat{\theta}_{n})$ on $g_{t}(X_{i},\widehat{\theta}_{n})$ that do not depend on the projection direction $\beta$. As such, $e_{i}^{pro,\ast}(t;\widehat{\theta}_{n})$ are nothing but the bootstrap counterparts of $e_{i}^{pro}(t;\widehat{\theta}_{n})$ in \eqref{e_proj}. 

The multiplier bootstrapped test statistic $CvM_{n}^{dpro,\ast}$ has attractive theoretical and empirical
properties. First, it does not require computing new parameter estimates at each bootstrap draw, reducing the computational intensity of the proposed procedure. Second, thanks to the use of the second projection in $\mathcal{P}_{n,t}1\left(X^\top_{i}\beta\leq u\right)$, its implementation does not require using estimators that admit an asymptotically linear representation and thus allows for a wider range of estimators. Third, thanks to the closed-form representation in (\ref{cvm_closed}), $A_{ijr}$ does not need to be computed for each bootstrap sample. Finally, it does not involve any tuning parameters such as bandwidths. All these features greatly alleviate the computational cost of our testing procedure. 

To establish the asymptotic validity of the proposed multiplier bootstrap procedure described in Steps 1--5 above, let
\[
R_{n,t}^{dpro,\ast}(\beta,u;\widehat{\theta}_{n})=\frac{1}{\sqrt{n}}\sum_{i=1}^{n}V_{i}\,e_{i}(t;\widehat{\theta}_{n})\mathcal{P}_{n,t}1\left(X^\top_{i}\beta\leq u\right)
\]
denote the multiplier bootstrapped version of $R_{n,t}^{dpro}(\beta,u;\widehat{\theta}_{n})$, with the sequence of multiplies $\{V_{i}\}_{i=1}^n$ as described in Step 1. In addition, the multiplier bootstrapped version of $CvM_n^{drpo}$ is 
\begin{align*}
CvM_{n}^{dpro,\ast}=&\sum_{t\in\mathcal{T}}a_{n}(t)\int_{\Pi_{pro}}\left(R_{n,t}^{dpro,\ast}(\beta,u;\widehat{\theta}_{n})\right)  ^{2}\,F_{n,\beta}(du)\,d\beta,
\end{align*}
whose closed-form expression is given by \eqref{cvm_closed_boot}. It is shown in Appendix A that
\[
R_{n,t}^{dpro,\ast}(\beta,u;\widehat{\theta}_{n})=\frac{1}{\sqrt{n}}\sum
_{i=1}^{n}V_{i}\,e_{i}(t;\theta_{0})\mathcal{P}_{t}1\left(X^\top_{i}\beta\leq u\right)+o_p(1),
\]
uniformly in $(\beta^\top,u)^\top$ for each $t\in\mathcal T$. Thus, $R_{n,t}^{dpro,\ast}(\beta,u;\widehat{\theta}_{n})$ is asymptotically invariant to $\widehat\theta_n$, and this holds under $H_{0}$, $H_{1}$, and $H_{1,n}$. The next theorem formally states the asymptotic validity of the multiplier bootstrap procedure.

\begin{theorem}
\label{th:boot_expansion} Suppose Assumptions \ref{A1}-\ref{A4} hold. Then, we have $CvM_{n}^{dpro,\ast}\underset{\ast}{\overset{d}{\rightarrow}}CvM_{\infty}^{dpro}$ in probability under the bootstrap law, where $CvM_{\infty}^{dpro}$ is the same distribution as defined in Theorem \ref{theorem1}, and $\underset{\ast}{\overset{d}{\rightarrow}}$ denotes weak convergence under the bootstrap law, i.e., conditional on the original sample $\left\{  (T_{i},X_{i}^{\top})^{\top}\right\}  _{i=1}^{n}$.
\end{theorem}

Theorem \ref{th:boot_expansion} states that the bootstrapped test statistic $CvM_{n}^{dpro,\ast}$ converges to the null limiting distribution of $CvM_{n}^{dpro}$ conditional on the original sample. The fact that $CvM_{n}^{dpro,\ast}$ has the same limiting distribution under $H_{0}$, $H_{1}$, and $H_{1,n}$ is what allows the proposed multiplier bootstrap procedure to work.

\begin{remark}\label{rem:general}
Our proposed double-projection procedure can be readily used to test the validity of conditional moment restrictions $\mathbb{E}\left[  e(t;\theta_0)|X\right]=0~a.s.$ for all $t\in\mathcal{T}$, for an appropriately defined generalized error $e(t;\theta_0)$. This includes classical regression models, where $\mathcal{T}=\{1\}$ the outcome of interest in $Y$, the parametric model for $\mathbb{E}[Y|X]$ is $q(X,\theta)$, and $e(1;\theta)=Y-q(X,\theta)$. One can also test conditional mean and variance models in the spirit of \citet{Escanciano2008}, where $\mathcal{T}=\{1,2\}$, $e(1;\theta)=Y-q_1(X,\theta_1)$, $e(2;\theta)=(Y-q_1(X,\theta_1))^2 - q_2(X,\theta_2)$, with $q_1(X,\theta_1)$ being a parametric model for $\mathbb{E}[Y|X]$, and $q_2(X,\theta_2)$ a parametric model for the conditional variance of $Y-\mathbb{E}[Y|X]$. In these cases, all one needs to do is adjust the definition of the errors and score functions. Importantly, we do not require that parametric errors (and the outcome variables) to be continuous.

We also stress that it is straightforward to extend our proposal to test even more general conditional moment restrictions of the following form: 
\begin{align*}
H_{0}^{cmr} &  :\mathbb{P}\left(  \mathbb E[\rho(Z,\theta_0)|X]=0\right)=1\,\,\,\text{for some}\,\,\,\theta_{0}\in\Theta\subset\mathbb{R}^{d_\theta},\\
H_{1}^{cmr} &  :\mathbb{P}\left(  \mathbb E[\rho(Z,\theta)|X]=0\right)  <1\,\,\,\text{for any}\,\,\,\theta\in\Theta\subset\mathbb{R}^{d_\theta},
\end{align*}
where $\rho(z,\theta)$ is a known function, not necessarily differentiable with respect to $\theta$; see, e.g., \cite{whang2001consistent}, \cite{delgado2006consistent}, and \cite{escanciano2010approximating}. A special case is when $Z=(Y,X)$, $\rho(Z,\theta)=Y-q(X,\theta)$ for some parametric function $q(X,\theta)$, and $Y$ is the dependent variable, as we just discussed above. Another interesting case is when $\rho(Z,\theta)=1(Y-q(X,\theta)\leq 0)-\tau$ with $\tau\in(0,1)$; then $q(X,\theta_0)$ is the $\tau$-quantile of the conditional distribution of $Y$ given $X$. For the general case, however, the dependent variable may be related to the regressors through the implicit function $\rho(\cdot,\cdot)$. Then, by letting $e(\theta)=\rho(Z,\theta)$ and $g(X,\theta)=\partial\mathbb E[\rho(Z,\theta)|X]/\partial\theta$, we can straightforwardly generalize our double-projection methodology to test $H_{0}^{cmr}$ against $H_{1}^{cmr}$.
\end{remark}

\section{Specification tests for multiple-index models\label{extension_sim}}
In the previous sections, we proposed omnibus-type specification tests that aim to detect the inadequacy of general parametric models. In particular, the working parametric models do not assume any dimension-reducing structure under the null and/or under the alternative. In many applications, however, using models with single/multiple index structures is common. In such cases, getting some additional insights about potential reasons for rejecting the putative model is interesting.

There has been a large amount of interest in statistical inference in (single) index models. \cite{horowitz1994testing}, \cite{Fan1996}, \cite{fan1997simple}, \cite{ait2001goodness}, \cite{hardle2001bootstrap}, \cite{Stute2002}, \cite{xia2004goodness}, \cite{stute2005nonparametric}, \cite{xia2009model}, \cite{escanciano2010testing}, \cite{guo2016model} and \cite{maistre2019nonparametric}, among others, proposed various approaches to testing generalized linear models or testing parametric or semiparametric (single) index assumptions. In the following, we investigate three relevant testing problems in the framework of index-type GPS. In Section \ref{extension_sim1}, we briefly describe a method for testing an index model (IM hereafter) with a known link function against a nonparametric alternative (i.e., not restricting ourselves to an IM with a known link function). To further uncover the potential sources of misspecification for the assumed IM, in Section \ref{extension_sim2}, we discuss the problem of directionally testing an IM with a known link function against a semiparametric alternative of IM with an unknown link function. Section \ref{extension_sim3} outlines how to test a semiparametric single-index model against a nonparametric alternative. We emphasize that the test we discuss in Section \ref{extension_sim3} is very different from all other tests discussed in the paper as the null model is semiparametric, i.e., it involves an infinite-dimensional nuisance function. 

\subsection{Testing a parametric IM against a nonparametric alternative\label{extension_sim1}}
For practical convenience, researchers often impose some dimension-reducing structure on the GPS. When the treatment is binary or ordered multinomial, a popular choice is the class of single-index models, namely, $q_t(X,\theta)=q_{t}(X^\top\theta)$, where $X^\top\theta$ is the single linear-index, and $q_{t}(\cdot): \mathbb R\mapsto [0,1]$ is a known link function for each $t\in\mathcal T$ (e.g., the multinomial logit link with $q_{t}(\cdot)$ specified as the cumulative logistic distribution function or the probit link with $q_{t}(\cdot)$ specified as the cumulative normal distribution function).\footnote{The multinomial ordered discrete choice model has an intercept that varies with $t$.} With unordered multinomial  choices, one usually adopts a multiple index model, $q_t(X,\theta)=q_{t}(X^\top\theta_1,\dots,X^\top\theta_J)$ as in the multinomial logit model (\ref{eq:mult_logit}). Since the binary and ordered multinomial cases are special cases of the unordered multinomial one, we focus on the latter.

For a generic $\theta=(\theta_{1}^\top,\dots, \theta_{J}^\top)^\top$, let $\widetilde{X}_\theta = (X^\top\theta_{1},\dots,X^\top\theta_{J})$ denote the vector of linear indexes. In this section, the null hypothesis of interest is
\begin{equation}
H_{0}^{im}:\mathbb{P}\left(  \mathbb{E}\left[  e^{im}(t;\theta_0)|X\right]=0\right)  =1\text{ for some }\theta_0\in\Theta\subset\mathbb{R}^{d_\theta}\text{ and all }t\in\mathcal{T}, \label{sim0}%
\end{equation}
where $\theta_0=(\theta_{0,1}^\top,\dots, \theta_{0,J}^\top)^\top$, and $e^{im}(t;\theta)\equiv 1(T=t)-q_{t}(\widetilde{X}_{\theta})$. The alternative hypothesis $H_{1}^{im}$ is the negation of $H_0^{im}$. This is an omnibus test of a multiple-index parametric model against a nonparametric alternative.

Testing $H_{0}^{im}$ against $H_{1}^{im}$ is equivalent to testing \eqref{cmr} against \eqref{cmr_alter} but with $q_t(X,\theta)=q_{t}(\widetilde{X}_{\theta})$ and the generalized error $e(t;\theta)$ replaced by $e^{im}(t;\theta)\equiv 1(T=t)-q_{t}(\widetilde{X}_{\theta})$. As such, we can consider the following generalized residual marked double-projected empirical process as a special case of $R^{dpro}_{n,t}(\beta,u;\widehat\theta_n)$, 
\begin{equation*}
\frac{1}{\sqrt n}\sum_{i=1}^ne_i^{im}(t;\widehat\theta_n)\mathcal P^{im}_{n,t}1(X_i^\top\beta\leq u), \quad \left(\beta^\top,u\right)^\top\in\Pi_{pro}, \label{sim_special_case}
\end{equation*}
where $e_i^{im}(t;\widehat\theta_n)\equiv 1(T_i=t)-q_{t}(\widetilde{X}_{i,\widehat{\theta}_n})$ is the generalized residual under the multiple index model and the projection operator $\mathcal P^{im}_{n,t}1(X_i^\top\beta\leq u)$ is as defined in \eqref{dp_weight} but with the score function $g_{t}(x,\theta)=\partial q_{t}(\widetilde{x}_{{\theta}})/\partial\theta$ for each $t\in\mathcal{T}$. The associated theoretical results are omitted as they are the same as those in previous sections.

\subsection{Testing a parametric IM against a semiparametric alternative\label{extension_sim2}}

In this section, we discuss how one can directionally test a parametric index model against a semiparametric index model, where the semiparametric component comes from link functions being unknown. Formally, we want to test the null hypothesis
\begin{equation}
H_{0}^{im1}: \mathbb{P}\left(  \mathbb{E}\left[  e^{im}(t;\theta_0)|\widetilde{X}_{\theta_0}\right]=0\right)  =1\text{ for some }\theta_0\in\Theta\subset\mathbb{R}^{d_\theta}\text{ and all }t\in\mathcal{T}, \label{sim01}%
\end{equation}
against the directional alternative
\begin{equation}
H_{1}^{im1}: \mathbb{P}\left( \mathbb{E}\left[1(T=t) - \mu_t(\widetilde{X}_{\theta})|\widetilde{X}_{\theta}\right]=0\right)  =1 \text{ for any}\,\,\,\theta\in\Theta\subset\mathbb R^{d_\theta}\,\,\,\text{and/or some}\,\,\, t\in\mathcal T,  \label{sim11}
\end{equation}
where, $\mu_t(\cdot): \mathbb R^J\mapsto [0,1]$ is an unknown link function such that $\mathbb P\left(\mu_t(\widetilde{X}_{\theta})=q_{t}(\widetilde{X}_{\theta})\right)<1$ for any $\theta\in\Theta$ and/or some $t\in\mathcal T$. 

\cite{horowitz1994testing} and \cite{fan1997simple} considered similar testing problems specialized to the single-index setup and constructed their test statistics by comparing a nonparametric estimate of $\mathbb P(T=t|X^\top\widehat\theta_n=v)$ with $q_{t}(v)$. We pursue a different route by using the integrated moment approach with double-projections and allowing for multiple-index models.

The key insight we provide here is to re-express the null hypothesis (\ref{sim01}) as
\begin{align}
H_{0}^{im1}:\mathbb{E}\left[  e^{im}(t;\theta_0)1\left(\widetilde{X}_{\theta_0}^\top\beta\leq u\right)  \right] =0&\text{ a.e. }\left(\beta^{\top},u\right)  ^{\top}\in\Pi_{pro}^{im},\notag\\
&\text{for some }\,\theta_0\in\Theta\subset\mathbb{R}^{d_\theta}\text{ and all }t\in\mathcal{T},  
\end{align}
where
 $\Pi_{pro}^{im}\equiv\mathbb{S}^{J}\times\overline{\mathbb R}$ denotes the projected space under the index model.  In light of our previous discussions, this characterization of the null immediately suggests using the following generalized residual marked double-projected empirical process with \emph{estimated} multiple indexes,
\begin{equation*}
M^{dpro}_{n,t}(\beta, u;\widehat\theta_n)\equiv\frac{1}{\sqrt n}\sum_{i=1}^ne_i^{im}(t;\widehat\theta_n)\mathcal P_{n,t}^{im}1(\widetilde{X}_{i,\widehat{\theta}_n}^{\top}\beta \leq u), \quad\left(  \beta^{\top},u\right)  ^{\top}\in\Pi_{pro}^{im},
\end{equation*}
where
\begin{equation}
\mathcal P_{n,t}^{im}1(\widetilde{X}_{i,\widehat{\theta}_n}^{\top}\beta \leq u)  \equiv 1(\widetilde{X}_{i,\widehat{\theta}_n}^{\top}\beta \leq u)-g_{t}(X_i,\widehat{\theta}_n)^\top\Delta_{n,t}^{-1}(\widehat{\theta}_{n})G_{n,t}^{im}(\beta, u;\widehat{\theta}_{n}), \label{new_sim_2}
\end{equation}
with $g_{t}(X_i,\widehat{\theta}_n)$ and $\Delta_{n,t}(\widehat{\theta}_{n})$ defined as before, and
\[
G_{n,t}^{im}(\beta,u;\widehat{\theta}_{n})=\frac{1}{n}\sum_{i=1}^{n}g_{t}(X_i,\widehat{\theta}_n)1(\widetilde{X}_{i,\widehat{\theta}_n}^{\top}\beta \leq u).
\]

Based on $M^{dpro}_{n,t}(\beta, u;\widehat\theta_n)$, we can use the following class of $CvM$-type test statistics to test $H_{0}^{im1}$ against $H_{1}^{im1}$:
\begin{align}
CvM_n^{im,dpro}=&\sum_{t\in\mathcal{T}}a_{n}(t)\int_{\Pi_{pro}^{im}}\left(M_{n,t}^{dpro}(\beta,u;\widehat{\theta}_{n})\right)^{2}\,F^{im}_{n,\beta,{\widehat{\theta}_n}}(du)\,d\beta, \label{cvm_sim2}
\end{align}
where $F_{n,\beta,{\widehat{\theta}_n}}^{im}(u)=n^{-1}\sum_{i=1}^{n}1\left(\widetilde{X}_{i, \widehat{\theta}_n}^\top\beta\leq u\right)  $ is the empirical distribution function of the one-dimensional projected estimated indexes $\left\{\widetilde{X}_{i, \widehat{\theta}_n}^\top\beta\right\}_{i=1}^{n}$ for any fixed projected direction $\beta\in\mathbb{S}^{J}$, and $d\beta$ is the rescaled uniform density on the unit sphere $\mathbb{S}^{J}$. 

Note that $CvM_n^{im,dpro}$ in (\ref{cvm_sim2}) resembles $CvM_n^{dpro}$ in (\ref{cvm}), with the caveat that now we use the vector of estimated linear indexes $\left\{\widetilde{X}_{i,\widehat{\theta}_n}\right\}_{i=1}^n$, instead of the observed covariates $\left\{X_i\right\}_{i=1}^n$ in the projection steps. As this difference changes the computational aspect of the test statistic and also its asymptotic properties, we discuss these differences in the following.

In terms of computation, by building on Lemma \ref{lemma1} we can show that $CvM_n^{im,dpro}$ in (\ref{cvm_sim2}) can be expressed in closed-form as
\begin{equation}
CvM_n^{im,dpro}= \sum_{t\in\mathcal{T}}a_{n}(t)\frac{1}{n^{2}}\sum_{i=1}^{n}\sum_{j=1}^{n}\sum_{r=1}^{n}e_{i}^{im,pro}(t;\widehat{\theta}_{n})e_{j}^{im,pro}(t;\widehat{\theta}_{n})A_{ijr}^{im}, \label{cvm_closed_im}%
\end{equation}
where
\begin{equation}
e_{i}^{im,pro}(t;\widehat{\theta}_{n})=e_{i}^{im}(t;\widehat{\theta}_{n})-g_{t}(X_i,\widehat{\theta}_n)^\top\Delta_{n,t}^{-1}(\widehat{\theta}_{n})\frac{1}{n}\sum_{s=1}^{n}g_{t}(X_s,\widehat{\theta}_n) e_{s}^{im}(t;\widehat{\theta}_{n}),\quad i=1,\ldots,n,\label{e_im,pro}
\end{equation}
and $A_{ijr}^{im } = A_{ijr}^{im(0)}\times\left(\pi^{J/2-1}\right)/\Gamma\left(J/2\right)$ with
\begin{small}
\[
A_{ijr}^{im\left(  0\right)  }=\left\{
\begin{array}
[c]{ll}%
2\pi, ~~~~~~~~~~~~~~~~~~~~~~~~~~~~~~~~  if ~~\widetilde{X}_{i, \widehat{\theta}_n}=\widetilde{X}_{r, \widehat{\theta}_n}=\widetilde{X}_{j, \widehat{\theta}_n},\\
\pi, ~~~~~~~~~~~~~~~~~~~~~~~~~~~~~~~~ ~if ~~\widetilde{X}_{i, \widehat{\theta}_n}=\widetilde{X}_{j, \widehat{\theta}_n},~or~\widetilde{X}_{i, \widehat{\theta}_n}=\widetilde{X}_{r, \widehat{\theta}_n},~or~\widetilde{X}_{j, \widehat{\theta}_n}=\widetilde{X}_{r, \widehat{\theta}_n},\\
\pi-\arccos\left(  \dfrac{\left( \widetilde{X}_{i, \widehat{\theta}_n}-\widetilde{X}_{r, \widehat{\theta}_n}\right)  ^\top\left(
\widetilde{X}_{j, \widehat{\theta}_n}-\widetilde{X}_{r, \widehat{\theta}_n}\right)  }{\left\Vert \widetilde{X}_{i, \widehat{\theta}_n}-\widetilde{X}_{r, \widehat{\theta}_n}\right\Vert \left\Vert \widetilde{X}_{j, \widehat{\theta}_n}%
-\widetilde{X}_{r, \widehat{\theta}_n}\right\Vert }\right), ~~~~~~~~~~~~~ otherwise.
\end{array}
\right.
\]
\end{small}
This characterization helps us implement the multiplier bootstrap procedure and on other practical computational aspects without needing to compute a continuum of projections.

In terms of the statistical properties of our test statistic $CvM_n^{im,dpro}$, if we show that there is no estimation effect from using the estimated indexes in the weighting functions, we can resort to Theorem \ref{theorem1} to establish size control. Indeed, as we show in Appendix A, we have that, uniformly in $(\beta^\top,u)^\top \in \Pi_{pro}^{im}$ and for each $t\in\mathcal{T}$,
\begin{align*}
&M^{dpro}_{n,t}(\beta, u;\widehat\theta_n)\\
=&\frac{1}{\sqrt n}\sum_{i=1}^ne_i^{im}(t;\widehat\theta_n)\left(1(\widetilde{X}_{i,\theta_0}^{\top}\beta \leq u)-g_{t}(X_i,\widehat{\theta}_n)^\top\Delta_{n,t}^{-1}(\widehat{\theta}_{n})G_{n,t}^{im}(\beta, u;\widehat{\theta}_{n})\right)+ o_p(1)\\
=&\frac{1}{\sqrt n}\sum_{i=1}^ne_i^{im}(t;\theta_0)\left(1(\widetilde{X}_{i,\theta_0}^{\top}\beta \leq u)-g_{t}(X_i,\theta_0)^\top\Delta_{t}^{-1}(\theta_{0})G_{t}^{im}(\beta, u;\theta_{0})\right)+ o_p(1)\\
\equiv&\frac{1}{\sqrt n}\sum_{i=1}^ne_i^{im}(t;\theta_0)\mathcal P_t^{im}1(\widetilde{X}_{i,\theta_0}^{\top}\beta \leq u)+o_p(1),
\end{align*}
directly leading to the following result.

\begin{proposition}
\label{prop2} Suppose Assumptions \ref{A1}-\ref{A3} hold. Consider a multiple-index specification, $q_{t}(X,\theta) = q_{t}(\widetilde{X}_\theta)$ for all $t \in \mathcal{T}$, such that $q_{t}(\cdot)$ is known up to the finite-dimensional parameters $\theta$. Assume that the underlying distribution of the projected-multiple indexes, $F_{\beta,\theta_0}(\cdot)$, is absolutely continuous with respect to the Lebesgue measure. Then, under the null hypothesis $H_{0}^{im1}$ in (\ref{sim01}), for any sequence $a_{n}(t)=a(t)+o_{p}\left(  1\right)  $, with $a(t)>0$ and $0<\sum_{t\in\mathcal{T}}a(t)\leq C<\infty$, we have that
\[
CvM_n^{im,dpro}\xrightarrow{d}CvM_\infty^{im,dpro}\equiv\sum_{t\in\mathcal{T}}a(t)\int_{\Pi_{pro}^{im}}\left(  M^{dpro}_{\infty,t}(\beta,u)\right)  ^{2}\,F_{\beta,\theta_0}(du)d\beta  ,
\]
where $M^{dpro}_{\infty,t}$ is a Gaussian process with mean zero and covariance structure
\begin{equation}
\mathbb{K}_{t}^{im,dpro}\left(  \left(  \beta,u\right)  ,\left(  \beta^{\prime},u^{\prime}\right)  \right)  =\mathbb{E}\left[  \widetilde{\sigma}_{t}^{2}(\widetilde{X}_{{\theta}_0})\mathcal{P}_{t}^{im}1\left(\widetilde{X}_{\theta_0}^\top\beta\leq u\right)  \mathcal{P}_{t}^{im}1\left(\widetilde{X}_{\theta_0}^\top\beta'\leq u^{\prime}\right)  \right]  , \label{kw_im_dpro}%
\end{equation}
where  $\widetilde{\sigma}_{t}^{2}(\widetilde{X}_{{\theta}_0})=\mathbb{E}\left[  \varepsilon^{2}(t)|\widetilde{X}_{{\theta}_0}\right]=q_{t}(\widetilde{X}_{{\theta}_0})\left(1-q_{t}(\widetilde{X}_{{\theta}_0})\right)  $ is the conditional variance function of the generalized error $\varepsilon(t)$ given $\widetilde{X}_{{\theta}_0}$ under the null $H_0^{im1}$.
\end{proposition}

In order to study the power properties of our proposed test, recall that $p_t(x)=\mathbb P(T=t|X=x)$ is the true, unknown GPS, which does not have to satisfy the semiparametric index restriction stated in the alternative hypothesis $H_1^{im1}$. To proceed with the analysis of the asymptotic global power properties of $CvM_n^{im,dpro}$, we can show that, uniformly in $(\beta^\top,u)^\top \in \Pi_{pro}^{im}$, 
\begin{equation*}
\frac{M^{dpro}_{n,t}(\beta,u;\widehat\theta_n)}{\sqrt n}\xrightarrow{p}\mathbb{E}\left[  \left(  p_{t}\left(  X\right)-q_{t}(\widetilde{X}_{{\theta}^*})  \right)  \mathcal{P}_{t}^{im}1\left(\tilde{X}_{{\theta}^*}^\top\beta\leq u\right)  \right],
\end{equation*}
for each $t\in\mathcal T$ under $H_1^{im1}$, where, using obvious notation, $\mathcal{P}_{t}^{im}1\left(\widetilde{X}_{{\theta}^*}^\top\beta\leq u\right)$ is the probability limit of \eqref{new_sim_2}.

If we ignore the less interesting class of alternative hypotheses under which the difference between $q_t(\widetilde{X}_{{\theta}^*})$ and $\mu_{t}(\widetilde{X}_{{\theta}^*})$ is collinear with the score function $g_{t}(X,\theta^*)$, we have that
\begin{align*}
&\mathbb E\left[\left(p_t(X)-q_{t}(\widetilde{X}_{{\theta}^*})\right)\mathcal P_{t}^{im}1(\widetilde{X}_{{\theta}^*}^\top\beta\leq u)\right]\\
=&\mathbb E\left[\left(p_t(X)-\mu_{t}(\widetilde{X}_{{\theta}^*})\right)\mathcal P_{t}^{im}1(\widetilde{X}_{{\theta}^*}^\top\beta\leq u)\right]\\
&+\mathbb E\left[\left(\mu_t(\widetilde{X}_{{\theta}^*})-q_{t}(\widetilde{X}_{{\theta}^*})\right)\mathcal P_{t}^{im}1(\widetilde{X}_{{\theta}^*}^\top\beta\leq u)\right].
\end{align*}
On one hand, when $p_t(x)=\mu_{t}(\tilde{x}_{{\theta}^*})$ holds (i.e., the true GPS satisfies the semiparametric IM assumption under $H_1^{im1}$), the fact that $q_t(v)\neq \mu_{t}(v)$ would be able to guarantee $\mathbb E\left[\left(\mu_t(\widetilde{X}_{{\theta}^*})-q_{t}(\widetilde{X}_{{\theta}^*})\right)\mathcal P_{t}^{im}1(\widetilde{X}_{{\theta}^*}^\top\beta\leq u)\right]\neq 0$, thus implying that our test is consistent against $H_1^{im1}$ in \eqref{sim11}. 

On the other hand, if the true model does not satisfy an index restriction, i.e., $p_t(x)\neq \mu_{t}(\tilde{x}_{{\theta}^*})$, it is possible that $\mathbb{E}\left[\left(p_t(X)-q_{t}(\widetilde{X}_{{\theta}^*})\right)\mathcal P_{t}^{im}1(\widetilde{X}_{{\theta}^*}^\top\beta\leq u)\right]=0$, even if the null hypothesis $H_0^{im1}$ is false. As so, tests based on $M^{dpro}_{n,t}(\beta,u;\widehat\theta_n)$ are inconsistent against the more general class of alternative hypotheses consisting of the negation of \eqref{sim01}. These findings, however, are typical features of directional-type tests; see, e.g., \cite{horowitz1994testing}. It is also straightforward to show that our test $CvM_{n}^{im,dpro}$ is able to detect a broad range of local alternatives, similar to the discussion in Theorem \ref{thh1n}. We omit the details to avoid repetition.

In practice, one can leverage the closed-form representation of our test statistic in (\ref{cvm_closed_im}) to compute critical values and/or $p$-values with the assistance of a convenient multiplier bootstrap procedure. Similar to Theorem \ref{th:boot_expansion}, we can show that to compute the bootstrap analog of $CvM_n^{im,dpro}$, all one needs to do is replace $e_{i}^{im,pro}(t;\widehat{\theta}_{n})$ in (\ref{cvm_closed_im}) with its ``perturbed''  version, 
\begin{equation}
e_{i}^{im,pro,*}(t;\widehat{\theta}_{n})=e_{i}^{im,*}(t;\widehat{\theta}_{n})-g_{t}(X_i,\widehat{\theta}_n)^\top\Delta_{n,t}^{-1}(\widehat{\theta}_{n})\frac{1}{n}\sum_{s=1}^{n}g_{t}(X_s,\widehat{\theta}_n) e_{s}^{im,*}(t;\widehat{\theta}_{n}),\label{e_im_boot}
\end{equation}
where $e_{i}^{im,\ast}(t;\widehat{\theta}_{n})=V_{i}\,e_{i}^{im}(t;\widehat{\theta}_{n})$, $i=1,\dots,n$, and $\left\{V_{i},i=1,\ldots,n\right\}$ is a sequence of i.i.d. random variables with mean zero, variance one, and finite third moment. Given our results in Theorem \ref{th:boot_expansion}, it is easy to show that the multiplier bootstrap analog of $CvM_n^{im,dpro}$ converges to the null distribution of $CvM_{n}^{im,dpro}$ conditional on the original sample, establishing its asymptotic validity. We omit the details to avoid repetition of arguments.

\subsection{Testing a semiparametric single-index model against a nonparametric alternative\label{extension_sim3}}

So far, we have focused our attention on testing parametric models against omnibus or directional alternatives. Suppose that a researcher rejects a putative parametric IM, and then decides to use a more flexible semiparametric model while still maintaining the index restriction. By doing so, the researcher can still alleviate the ``curse of dimensionality''.  In this context, however, a natural question arises: is the semiparametric index model correctly specified? In this section, we describe how one can use a variant of our double-projection procedure to answer this type of question. Given that the setup here is very different from the rest of the paper, we restrict our attention to single-index models (SIM). Extensions for multiple-index models are possible but involve much more tedious arguments and steps.

More precisely, we are now interested in testing the following semiparametric-type null hypothesis
\begin{align}
H_{0}^{sim}: \,\,\,&\mathbb P\left(\mathbb E\left[e^{sim}(t;\theta_0, \mu_t)|X\right]=0\right)=1 \notag\\
&\text{for some}\,\,\,\theta_0\in\Theta\subset\mathbb R^{d_\theta}\,\,\,\text{and some}\,\,\,\mu_t(\cdot)\in\mathcal{C}_k,\,\,\text{and all}\,\,\, t\in\mathcal T,
\end{align}
where $e_i^{sim}(t;\theta_0, \mu_t)\equiv 1(T_i=t)-\mu_{t}(X_i^\top\theta_0)$, and $\mu_t(\cdot): \mathbb R\mapsto [0,1]$ is an unknown link function that belongs to $\mathcal{C}_k$, the space of real-valued, continuous (measurable) functions with bounded $k$-th derivative, $k>1$. The alternative hypothesis $H_{1}^{sim}$ is the negation of $H_{0}^{sim}$, i.e, 
\begin{align}
H_{1}^{sim}: \,\,\,&\mathbb P\left(\mathbb E\left[e^{sim}(t;\theta, \tilde{\mu}_t)|X\right]=0\right)<1,\notag\\
&\text{for any}\,\,\,\theta\in\Theta\subset\mathbb R^{d_\theta},\,\,\text{any } \tilde{\mu}_t(\cdot)\in\mathcal{C}_k,\,\,\,\text{and/or some}\,\,\, t\in\mathcal T.
\end{align}

Henceforth, we assume that the linear combination $X^\top\theta_0$ admits a probability density function (PDF) $f(\cdot)$. Let $K(v)$ be the univariate kernel function and $h$ be a bandwidth parameter that shrinks to zero at an appropriate rate as $n\rightarrow\infty$. In order to test the above semiparametric conditional moment restriction, in light of our previous discussion and \citet{Escanciano2006b},\footnote{We stress that \citet{Escanciano2006b} does not consider tests for semiparametric models like the one being discussed here.} it is natural first to consider the following density-weighted generalized residual marked projected empirical process:
\begin{align*}
S^{pro}_{n,t}(\beta,u;\widehat\theta_n, \widehat{\mu}_t)=\frac{1}{\sqrt n}\sum_{i=1}^n\widehat e_i^{sim}(t;\widehat\theta_n, \widehat{\mu}_t)\widehat f(X_i^\top\widehat\theta_n)1(X_i^\top\beta\leq u), \quad  \left(\beta^\top,u\right)^\top\in\Pi_{pro},
\end{align*}
where $\widehat\theta_n$ is any $\sqrt n$-consistent estimator for $\theta_0$, say, the minimum average conditional variance estimator of \citet{Xia2002} and \citet{Xia2006a}, $\widehat e_i^{sim}(t;\widehat\theta_n, \widehat{\mu}_t)\equiv 1(T_i=t)-\widehat{\mu}_{t}(X_i^\top\widehat\theta_n)$ is the semiparametric generalized residual, 
\begin{equation*}
\widehat{\mu}_t(X_i^\top\widehat\theta_n)=\frac{\frac{1}{n-1}\sum_{j\neq i}^n\frac{1}{h}K\left(\frac{(X_i-X_j)^\top\widehat\theta_n}{h}\right)1(T_j=t)}{\widehat f(X_i^\top\widehat\theta_n)}
\end{equation*}
and
\begin{equation*}
\widehat f(X_i^\top\widehat\theta_n)=\frac{1}{n-1}\sum_{j\neq i}^n\frac{1}{h}K\left(\frac{(X_i-X_j)^\top\widehat\theta_n}{h}\right)
\end{equation*}
are the standard leave-one-out kernel estimates for single-index model $\mu_{t}(X_i^\top\theta_0)=\mathbb E[1(T_i=t)|X_i^\top\theta_0]$ and the PDF $f(X_i^\top\theta_0)$, respectively.\footnote{The kernel density estimate $\widehat f(X_i^\top\widehat\theta_n)$ is employed in $S^{pro}_{n,t}(\beta,u;\widehat\theta_n,\widehat{\mu}_t)$ to avoid random denominators and ease theoretical derivations; see, e.g., \cite{powell1989semiparametric} and \cite{Delgado&GM2001}.} Density-weighted statistics like $S^{pro}_{n,t}(\beta,u;\widehat\theta_n,\widehat{\mu}_t)$ avoid random denominators, which simplifies the large sample derivations and also usually translates to better finite sample properties.

Clearly, $S^{pro}_{n,t}(\beta,u;\widehat\theta_n,\widehat{\mu}_t)$ satisfies the following decomposition:
\begin{align*}
S^{pro}_{n,t}(\beta,u;\widehat\theta_n,\widehat{\mu}_t)=&\frac{1}{\sqrt n}\sum_{i=1}^n\left(1(T_i=t)-\mu_t(X_i^\top\widehat\theta_n)\right)\widehat f(X_i^\top\widehat\theta_n)1(X_i^\top\beta\leq u)\\
&-\frac{1}{\sqrt n}\sum_{i=1}^n\left(\widehat \mu_t(X_i^\top\widehat\theta_n)-\mu_t(X_i^\top\widehat\theta_n)\right)\widehat f(X_i^\top\widehat\theta_n)1(X_i^\top\beta\leq u)\\
\equiv&S^{pro}_{n,t1}(\beta,u;\widehat\theta_n,{\mu}_t)-S^{pro}_{n,t2}(\beta,u;\widehat\theta_n,\widehat{\mu}_t),
\end{align*}
where $S^{pro}_{n,t1}(\beta,u;\widehat\theta_n,{\mu}_t)$ is affected by the finite-dimensional estimator of $\widehat\theta_n$ but not by the infinite-dimensional estimator $\widehat{\mu}_t$, whereas $S^{pro}_{n,t2}(\beta,u;\widehat\theta_n,\widehat{\mu}_t)$ is affected by both. By working with each of these terms separately, we show in Appendix A that, under $H_0^{sim}$, uniformly in $(\beta^\top,u)^\top\in\Pi_{pro}$, 
\begin{align}
&S^{pro}_{n,t}(\beta,u;\widehat\theta_n, \widehat{\mu}_t)\notag\\
=&\frac{1}{\sqrt n}\sum_{i=1}^n e_i^{sim}(t;\theta_0, {\mu}_t)f(X_i^\top\theta_0)\left(1(X_i^\top\beta\leq u)-\mathbb E[1(X_i^\top\beta\leq u)|X_i^\top\theta_0]\right)\notag\\
&-\sqrt n(\widehat\theta_n-\theta_0)^\top\mathbb E\left[\mu_t'(X^\top\theta_0)f(X^\top\theta_0)\left(X-\mathbb E[X|X^\top\theta_0]\right)1(X^\top\beta\leq u)\right]+o_p(1), \label{last_decom}
\end{align}
where $\mu_t'(v)=d\mu_t(v)/dv$ is the (unknown) first derivative of $\mu_t(v)$ for each $t\in\mathcal T$. It is interesting to remark that the quantity 
\begin{equation}
\sqrt n(\widehat\theta_n-\theta_0)^\top\mathbb E\left[\mu_t'(X^\top\theta_0)f(X^\top\theta_0)\left(X-\mathbb E[X|X^\top\theta_0]\right)1(X^\top\beta\leq u)\right] \label{1111}
\end{equation}
can be regarded as the \emph{total} or \emph{overall} parametric-type estimation effect due to using $\widehat\theta_n$ to estimate $\theta_0$, while the quantity 
\begin{equation}
\frac{1}{\sqrt n}\sum_{i=1}^n e_i^{sim}(t;\theta_0, {\mu}_t)f(X_i^\top\theta_0)\mathbb E\left[1(X_i^\top\beta\leq u)|X_i^\top\theta_0\right] \label{2222}
\end{equation}
can be regarded as the \emph{nonparametric}-type estimation effect due to using the nonparametric estimator $\widehat \mu_t(v)$ to replace $\mu_t(v)$ if the linear-index $v=x^\top\theta_0$ were known.

To formally establish \eqref{last_decom}, we need the following assumptions, on top of Assumptions \ref{A1} and \ref{A3}.

\begin{assumption}\label{Ass:smoothness}
The semiparametric propensity score model $\mu_{t}(v)$ is $k$-times continuously differentiable in $v$ with uniformly bounded derivatives, i.e., $\sup_v|d^jq_t(v)/dv^j|<C_j$ with positive constant $C_j$ for $j=1,\ldots, k$, $k\ge 1$.
\end{assumption}

\begin{assumption}\label{ass:kernel}
The kernel function $K(v)$ is bounded, symmetric around zero, and $M$-times continuously differentiable in $v$ with uniformly bounded derivatives ($M\geq 5$ is an integer). In addition, $K(v)$ satisfies $\int K(v)dv=1$, $\int v^jK(v)=0$ for $j=1,\ldots,k-1$, $\int v^k K(v)\neq 0$, $\int K^2(v)dv<\infty$, $\int |v|^k|K(v)|dv<\infty$, and $|v|^k|K(v)|\rightarrow 0$ as $|v|\rightarrow\infty$.
\end{assumption}

\begin{assumption}\label{ass:bandwidth}
The positive bandwidth sequence $h=h_n$ satisfies $h\rightarrow 0$, $nh^3\rightarrow\infty$ and $nh^{2k}\rightarrow 0$ as $n\rightarrow\infty$.

\end{assumption}

Assumption \ref{Ass:smoothness} assumes the link function is smooth, while Assumption \ref{ass:kernel} requires that $K(v)$ is a higher-order kernel. These are common assumptions in the literature on nonparametric kernel estimation. Assumption \ref{ass:bandwidth} is a standard assumption on the bandwidth that balances between biases and variances of $U$-processes, which appear in the proof. In particular, $nh^3\rightarrow\infty$ is a condition used in proving that the second order terms of $U$-processes are asymptotically uniformly negligible,  while the condition $nh^{2k}\rightarrow 0$ is used to control the bias terms. Note that if the standard normal kernel is used, it is differentiable to any order and $k=2$. Although our assumptions may be stronger than necessary, they make the proofs much easier.

Given the linear representation in \eqref{last_decom}, it is straightforward to establish the asymptotic null distribution of $S^{pro}_{n,t}(\beta,u;\widehat\theta_n, \widehat{\mu}_t)$ when $\sqrt{n}(\widehat{\theta}_{n}-\theta_0)$ admits an asymptotically linear representation. Even in that case, it is clear that its asymptotic null distribution would depend on how $\widehat\theta_n$ is obtained. Based on these more demanding and arguably less-desirable conditions, it is easy to build on \eqref{last_decom} to propose a $CvM$-type test, establish its power properties and use a multiplier-bootstrap akin to the one discussed in Section \ref{boot} to compute asymptotically-valid critical values; see, e.g., \cite{xia2004goodness} for a related proposal that relies on the standard indicator weight $1(X\leq \tilde{x})$ instead of the one-dimensional projected indicator weight $1(X^\top\beta\leq u)$.\footnote{Their bootstrap procedure also requires outcomes to be continuous, ruling out discrete response variables such as in the (generalized) propensity score models considered here.}

Instead of following this ``more-traditional'' path, one may propose a test insensitive to the choice of estimators for the nuisance parameters. In what follows, we argue that using a double-projection argument like those introduced in Section \ref{sec:dim_red}, we can eliminate the parametric-type estimation effect \eqref{1111}, but not the nonparametric-type of estimation effect \eqref{2222}. Still, using the double-projection procedure allows researchers to consider a wider variety of estimators for $\theta_0$ than if they were to propose test-statistics based on $S^{pro}_{n,t}(\beta,u;\widehat\theta_n, \widehat{\mu}_t)$.

To see this, let the (infeasible) double-projection weights be defined as
\begin{equation*}
\mathcal P_{t}^{sim}1(X_i^\top\beta\leq u)\equiv 1(X_i^\top\beta\leq u)-\mu_t'(X_i^\top\theta_0)f(X_i^\top\theta_0)\eta(X_i,\theta_0)^\top\Delta_t^{-1}(\theta_0)G_t(\beta,u;\theta_0),
\end{equation*}
where $\eta(x,\theta)=x-\varphi (x^\top\theta)$ with $\varphi(v)=\mathbb E[X|X^\top\theta=v]$,
\begin{equation*}
\Delta_t(\theta)=\mathbb E[\mu_t'(X^\top\theta)^2f(X^\top\theta)^2\eta(X,\theta)\eta(X,\theta)^\top],
\end{equation*}
and
\begin{equation*}
G_t(\beta,u;\theta)=\mathbb E[\mu_t'(X^\top\theta)f(X^\top\theta)\eta(X,\theta)1(X^\top\beta\leq u)].
\end{equation*}
The feasible version of $\mathcal P_{t}^{sim}1(X_i^\top\beta\leq u)$ is given by
\begin{small}
\begin{equation}
\mathcal{P}_{n,t}^{sim}1(X_i^{\top}\beta\leq u)  \equiv 1(X_i^{\top}\beta\leq u)-\widehat \mu_{t}'(X_i^\top\widehat\theta_n)\widehat f(X_i^\top\widehat\theta_n)\widehat\eta(X_i,\widehat\theta_n)^\top\Delta_{n,t}^{-1}(\widehat{\theta}_{n})G_{n,t}(\beta,u;\widehat{\theta}_{n}), \label{new_sim_121}
\end{equation}
\end{small}
where
\begin{small}
\begin{align*}
\widehat \mu'_t(X_i^\top\widehat\theta_n)=\frac{\frac{1}{(n-1)^2h^3}\sum_{j\neq i}^n\sum_{j'\neq i}^nK\left(\frac{(X_i-X_j)^\top\widehat\theta_n}{h}\right)K^{(1)}\left(\frac{(X_i-X_{j'})^\top\widehat\theta_n}{h}\right)\left(1(T_{j'}=t)-1(T_{j}=t)\right)}{\widehat f(X_i^\top\widehat\theta_n)^2}
\end{align*}
\end{small}
is the leave-one-out kernel estimator for $\mu'_t(X_i^\top\theta_0)\equiv d\mu_t(v)/dv|_{v=X_i^\top\theta_0}$, $\widehat\eta(X_i,\widehat\theta_n)=X_i-\widehat\varphi(X_i^\top\widehat\theta_n)$ with $\widehat\varphi(X_i^\top\widehat\theta_n)$ the leave-one-out kernel estimator for $\varphi(X_i^\top\theta_0)$ is given by
\begin{equation*}
\widehat \varphi(X_i^\top\widehat\theta_n)=\frac{\frac{1}{(n-1)h}\sum_{j\neq i}^nK\left(\frac{(X_i-X_{j})^\top\widehat\theta_n}{h}\right)X_j}{\widehat f(X_i^\top\widehat\theta_n)},
\end{equation*}
and
\begin{equation*}
\Delta_{n,t}(\widehat{\theta}_{n})=\frac{1}{n}\sum_{i=1}^{n}\widehat \mu_{t}'(X_i^\top\widehat\theta_n)^2\widehat f(X_i^\top\widehat\theta_n)^2\widehat\eta(X_i,\widehat\theta_n)\widehat\eta(X_i,\widehat\theta_n)^\top,
\end{equation*}
\begin{equation*}
G_{n,t}(\beta,u;\widehat{\theta}_{n})=\frac{1}{n}\sum_{i=1}^{n}\widehat \mu_{t}'(X_i^\top\widehat\theta_n)\widehat f(X_i^\top\widehat\theta_n) \widehat\eta(X_i,\widehat\theta_n) 1(X_i^{\top}\beta\leq u).
\end{equation*}
 
Consider the following density-weighted generalized residual marked double-projected empirical process
 \begin{small}
\begin{equation*}
S^{dpro}_{n,t}(\beta,u;\widehat\theta_n,\widehat{\mu}_t)=\frac{1}{\sqrt n}\sum_{i=1}^n\widehat e_i^{sim}(t;\widehat\theta_n,\widehat{\mu}_t)\widehat f(X_i^\top\widehat\theta_n)\mathcal P_{n,t}^{sim}1(X_i^\top\beta\leq u), \quad  \left(\beta^\top,u\right)^\top\in\Pi_{pro}.
\end{equation*}
 \end{small}
Note that, to compute $\mathcal{P}_{n,t}^{sim}1(X_i^{\top}\beta\leq u)$, one needs to estimate several infinite-dimensional parameters, including the derivatives $\mu'_t(\cdot)$. This makes the computation of $S^{dpro}_{n,t}(\beta,u;\widehat\theta_n,\widehat{\mu}_t)$ substantially more challenging than all the other double-projection empirical process previously considered in this paper.

Following similar arguments as proving \eqref{last_decom}, it is possible to show that, under some additional regularity conditions, uniformly in $(\beta^\top,u)^\top$, and under $H_0^{sim}$,
 \begin{small}
\begin{align*}
&S^{dpro}_{n,t}(\beta,u;\widehat\theta_n,\widehat{\mu}_t)\\
=&\frac{1}{\sqrt n}\sum_{i=1}^ne_i^{sim}(t;\theta_0,{\mu}_t))f(X_i^\top\theta_0)\left(\mathcal P_{t}^{sim}1(X_i^\top\beta\leq u)-\mathbb E[\mathcal P_{t}^{sim}1(X^\top\beta\leq u)|X_i^\top\theta_0]\right)+o_p(1)\\
=&\frac{1}{\sqrt n}\sum_{i=1}^n e_i^{sim}(t;\theta_0,{\mu}_t))f(X_i^\top\theta_0)\left(\mathcal P_{t}^{sim}1(X_i^\top\beta\leq u)-\mathbb E[1(X^\top\beta\leq u)|X_i^\top\theta_0]\right)+o_p(1),
\end{align*}
 \end{small}
\noindent where the second step follows because $\mathbb E[\eta(X_i,\theta_0)|X_i^\top\theta_0]=0$ $a.s.$. The good news now is that we do not have to deal with the parametric estimation effect due to using $\widehat\theta_n$, and therefore can avoid requiring the more stringent condition that $\sqrt{n}(\widehat{\theta}_{n}-\theta_0)$ admits an asymptotically linear representation. On the other hand, the \emph{nonparametric}-type estimation effect given in \eqref{2222} is still present.

The fact that we do not eliminate the nonparametric-type estimation effect using our double projections leads to some complications that were not present before. For instance, note that we can compute the following $CvM$ test statistic in closed-form: 
\begin{align*}
CvM_n^{sim,dpro}=&\sum_{t\in\mathcal{T}}a_{n}(t)\int_{\Pi_{pro}}\left(S^{dpro}_{n,t}(\beta,u;\widehat{\theta}_{n},\widehat{\mu}_t)\right)^{2}\,F_{n,\beta}(du)\,d\beta\\
=&\sum_{t\in\mathcal T}a_n(t)\frac{1}{n^2}\sum_{i=1}^n\sum_{j=1}^n\sum_{r=1}^n\widehat e_i^{sim,pro}(t;\widehat\theta_n)\widehat e_j^{sim,pro}(t;\widehat\theta_n)A_{ijr}, 
\end{align*}
where
\begin{align*}
\widehat e_{i}^{sim,pro}(t;\widehat{\theta}_{n})=&\widehat e_{i}^{sim}(t;\widehat{\theta}_{n})\widehat f(X_i^\top\widehat\theta_n)-\widehat \mu'_{t}(X_i^\top\widehat\theta_n)\widehat f(X_i^\top\widehat\theta_n)\widehat\eta(X_i,\widehat\theta_n)^\top\Delta_{n,t}^{-1}(\widehat{\theta}_{n}) \widehat{Q}_{n,t}
\end{align*}
and $$\widehat{Q}_{n,t} = \frac{1}{n}\sum_{s=1}^{n}\widehat \mu'_{t}(X_s^\top\widehat\theta_n)\widehat f(X_s^\top\widehat\theta_n)^2\widehat\eta(X_s,\widehat\theta_n)\widehat e_{s}^{sim}(t;\widehat{\theta}_{n}).$$

However, because the underlying null distribution of $S^{dpro}_{n,t}(\beta,u;\widehat\theta_n,\widehat{\mu}_t)$ is not invariant to the nonparametric estimator for the link function $\widehat{\mu}_t$, we need to account for its estimation effect in the implementation of a multiplier-bootstrap procedure. That is, a $CvM$-type multiplier bootstrapped test statistic can be constructed as 
\begin{align*}
CvM_n^{sim,dpro,\ast}=&\sum_{t\in\mathcal{T}}a_{n}(t)\int_{\Pi_{pro}}\left(S^{dpro,\ast}_{n,t}(\beta,u;\widehat{\theta}_{n},\widehat{\mu}_t)\right)^{2}\,F_{n,\beta}(du)\,d\beta,
\end{align*}
where 
\begin{align*}
S^{dpro,\ast}_{n,t}(\beta,u;\widehat\theta_n,\widehat{\mu}_t)=\frac{1}{\sqrt n}\sum_{i=1}^nV_i\widehat e_i^{sim}(t;\widehat\theta_n)\widehat f(X_i^\top\widehat\theta_n)\left(\mathcal P_{n,t}^{sim}1(X_i^\top\beta\leq u)-\widehat\phi(\beta,u;X_i^\top\widehat\theta_n)\right),
\end{align*}
with $\{V_i\}_{i=1}^n$ a sequence of multipliers defined in Section \ref{boot} and $\widehat\phi(\beta,u;X_i^\top\widehat\theta_n)$ the leave-one-out kernel estimator for $\phi(\beta,u;X_i^\top\theta_0)\equiv\mathbb E[1(X^\top\beta\leq u)|X^\top\theta_0=v]$ given by
\begin{equation*}
\widehat\phi(\beta,u;X_i^\top\widehat\theta_n)=\frac{\frac{1}{(n-1)h}\sum_{j\neq i}^nK\left(\frac{(X_i-X_{j})^\top\widehat\theta_n}{h}\right)1(X_j^\top\beta\leq u)}{\widehat f(X_i^\top\widehat\theta_n)}.
\end{equation*}
Unfortunately, given that $S^{dpro,\ast}_{n,t}(\beta,u;\widehat\theta_n,\widehat{\mu}_t)$ depends on  $\phi(\beta,u;X_i^\top\theta_0)$, which is indexed by $\beta$ and $u$, it is not clear if one can get an easy-to-use closed-form expression for $CvM_n^{sim,dpro,\ast}$. This suggests that one would need to resort to numerical integration procedures to compute critical values for $CvM_n^{sim,dpro}$ based on $CvM_n^{sim,dpro,\ast}$. Although this is feasible, it also seems much more computationally challenging. Given that these topics are sufficiently different from the rest of the paper, we leave their theoretical justifications and detailed implementation guidelines for future research.

\section{Data illustration \label{empirical_section}}

In this section, we apply our tests to analyze the goodness-of-fit of different GPS models used to study the effect of maternal smoking on birth weight. The dataset, available at \url{http://www.stata-press.com/data/r13/cattaneo2.dta}, is the excerpt from \cite{Almond2005} and \cite{Cattaneo2010} previously used by \cite{Lee2017}. It consists of observations from white mothers in Pennsylvania in the USA; like \cite{Lee2017}, we further restrict our sample to white and non-Hispanic mothers, total $3,754$ observations. The treatment variable, $T$, is a multi-valued variable that is equal to $0$ if the mother does not smoke during the pregnancy, equal to $1$ if the mother smokes, on average, between one and five cigarettes a day during the pregnancy, equal to $2$ if the mother smokes, on average, between six and ten cigarettes a day during the pregnancy, and equal to $3$ if the mother smokes, on average, more than eleven cigarettes a day during the pregnancy. The set of pre-treatment covariates $X$ we use are
the mother's age, number of prenatal care visits, and indicator variables for alcohol consumption during pregnancy, first prenatal visit in the first
trimester, whether there was a previous birth where the newborn died, twelve years of education (complete high-school), and more than twelve years of
educations (some college). The outcome of interest is the infant's birth weight measured in grams.

We start our analysis by analyzing the effect of the mother being a smoker during the pregnancy ($T>0$) versus not smoking during the pregnancy ($T=0$).
Given the binary nature of the ``being a smoker'' treatment, we estimate the propensity score using a logistic regression model with linear predictors including all aforementioned covariates. We then apply our proposed specification test to assess the goodness-of-fit of this simple propensity score model, using $9,999$ bootstrap replications. Our procedure yields a $p$-value of $0.18$, suggesting that our proposed testing procedure does not find any evidence of model misspecification at the usual significance levels.

Next, we move our attention to analyzing the effect of maternal smoking intensity during pregnancy, $T$. Given that the treatment $T$ is clearly ordered, we estimate the GPS using an ordered logit regression model with all covariates entering the model in a linear fashion. Although natural, we note that the ordered logit model imposes important restrictions on the data such as a proportional odds restriction. In practice, however, such restrictions may be too rigid for a given application. Indeed, our proposed specification test with $9,999$ bootstrap replications yield a
$p$-value of $0.08$, suggesting that the ordered logit model is misspecified at the 10\% significance level.

A relatively straightforward way to relax the proportional odd restrictions inherited in the ordered logit model is to ignore that the treatment $T$ is ordered and estimate the GPS using a multinomial logit linear regression model. In contrast with the ordered logit model, the multinomial logit model does not impose that the regressors' coefficients are the same across different treatment levels. Our specification test with $9,999$ bootstrap replications yield a $p$-value of $0.72$, suggesting that the multinomial logit model is a more suitable model for maternal smoking intensity during pregnancy $T$ than the order logit model.

Table \ref{tab:app} shows the estimates of the causal effects of smoking on birth weight based on the inverse probability weighting estimators for binary and multi-valued treatments, $ATE_{n}$ as in (B.3) and $ATE_{n,j,\ell}$ as in (B.7) in the Appendix, respectively. Although qualitatively similar, one can notice some differences in the standard errors and confidence intervals between the treatment effects estimates based on misspecified ordered logit model and those based on the multinomial logit model for the GPS. This illustrates one of the potential pitfalls of model misspecification.

\begin{table}[ptb]
\caption{Results from the empirical illustration: point estimates, standard error by the bootstrap, and 95\% confidence interval}%
\label{tab:app}
\centering
\centering\begin{adjustbox}{ max width=1\linewidth, max totalheight=1\textheight, keepaspectratio}
\begin{threeparttable}
\begin{tabular}{@{}lccc@{}} \hline
\toprule
\noalign{\vskip 2mm} & {\phantom{abd} Point Estimate\phantom{abd}} & {\phantom{abd}Standard Error\phantom{abd}} & {95\% Confidence Interval} \\
\noalign{\vskip 2mm} \multicolumn{4}{l}{\small{(a) The causal effect of mother being a smoker during pregnancy on infant's birth weight}} \\
\noalign{\vskip -1mm}\multicolumn{4}{l}{ \small{\phantom{abd}Estimators based on a logit propensity score model}} \\
\noalign{\vskip 2mm} \phantom{abcd} $\mathbb{E}\left[  Y\left(  1\right)  -Y\left(  0\right)\right]$  & -270 & 28 & (-326, -215)  \\
\noalign{\vskip 3mm} \multicolumn{4}{l}{\small{(b) The causal effect of mother's smoking intensity during pregnancy on infant's birth weight}} \\
\noalign{\vskip -1mm}\multicolumn{4}{l}{ \small{\phantom{abd}Estimators based on an ordered logit GPS model}} \\
\noalign{\vskip 2mm} \phantom{abcd} $\mathbb{E}\left[  Y\left(  1\right)  -Y\left(  0\right)\right]$    & -279 & 59 & (-401, -164)   \\
\phantom{abcd} $\mathbb{E}\left[  Y\left(  2\right)  -Y\left(  0\right)\right]$  & -244 & 46 & (-338, -156) \\
\phantom{abcd} $\mathbb{E}\left[  Y\left(  3\right)  -Y\left(  0\right)\right]$   & -279 & 39 & (-357, -202)  \\
\noalign{\vskip 2mm} \multicolumn{4}{l}{(c) \small{The causal effect of mother's smoking intensity during pregnancy on infant's birth weight}} \\
\noalign{\vskip -1mm}\multicolumn{4}{l}{\small{\phantom{abd}Estimators based on a multinomial logit GPS model}} \\
\noalign{\vskip 2mm}\phantom{abcd} $\mathbb{E}\left[  Y\left(  1\right)  -Y\left(  0\right)
\right]$   & -250 & 49 & (-346, -154) \\
\phantom{abcd} $\mathbb{E}\left[  Y\left(  2\right)  -Y\left(  0\right)\right]$  & -241 & 49 & (-338, -148) \\
\phantom{abcd} $\mathbb{E}\left[  Y\left(  3\right)  -Y\left(  0\right)\right]$   & -267 & 42 & (-347, -184)  \\
\bottomrule
\end{tabular}    \begin{tablenotes}[para,flushleft]
\footnotesize{
Note: Standard errors are computed using the empirical bootstrap with $9,999$ draws. 95\% confidence intervals based on the percentile bootstrap with $9,999$ draws.  See the main text for further details.}
\end{tablenotes}
\end{threeparttable}
\end{adjustbox}
\end{table}

\section{Conclusions and directions for further research\label{conclusion}}

In this article, we proposed a new class of specification tests for GPS models based on novel double-projected weight functions. We have shown that using double projections helps ameliorate the ``curse of dimensionality'' and avoids the complications associated with ``parameter estimation uncertainty'' commonly encountered in specification testing. We have shown that our proposed test statistics can be written in closed form, and that one can use an easy-to-implement multiplier bootstrap procedure to compute critical values as accurately as desired. We have also extended our double-projection proposal to test parametric multiple-index or semiparametric single-index GPS. The simulation results (in Appendix B) and the empirical application highlight that our proposed tests can serve as a valuable diagnostic tool in the context of multi-valued treatment effects. 

We anticipate that one can extend our proposal to test whether putative parametric conditional distributions, distributional regressions, or linear/nonlinear quantile regressions are correctly specified; see, e.g., \cite{bierens2012integrated}, \cite{Rothe2013a} and \cite{Escanciano2014a}. Due to the lack of a closed-form expression of $CvM_n^{sim,dpro,\ast}$ at the end of Section \ref{extension_sim3}, another direction of future research could focus on developing alternative bootstrap methods to obtain critical values of tests that are designed to test the semiparametric single-index assumption against a general nonparametric alternative and are robust to both the dimensionality of covariates and the estimation of parametric/nonparametric-type nuisance parameters. We leave a detailed analysis of these interesting extensions for future research. 

\onehalfspacing{\footnotesize
\bibliographystyle{ET}
\bibliography{pscore}
}

\end{document}